\documentclass[showpacs,twocolumn,superscriptaddress]{revtex4}
\usepackage{doi}
\usepackage{hyperref}
\hypersetup{
  colorlinks=true,        
  linkcolor=blue,         
  citecolor=cyan,         
}

\usepackage{graphicx}
\usepackage{dcolumn}
\usepackage{bm}
\usepackage{color}
\usepackage{enumitem}
\usepackage{amsmath}
\usepackage{amssymb}
\usepackage{subcaption}

\begin{document}%

\title{Effect of magnetized plasma on shadow and gravitational lensing of a Reissner-Nordstr\"om black hole}

\author{Yovqochev Pahlavon}
\email{yovqochevp@gmail.com}
\affiliation{Ulugh Beg Astronomical Institute, Astronomy St. 33, Tashkent 100052, Uzbekistan} 

\author{Farruh Atamurotov}
\email{atamurotov@yahoo.com}
\affiliation{New Uzbekistan University, Movarounnahr street 1, Tashkent 100000, Uzbekistan}
\affiliation{Institute of Theoretical Physics, National University of Uzbekistan, Tashkent 100174, Uzbekistan}
\affiliation{Institute of Fundamental and Applied Research, National Research University TIIAME, Kori Niyoziy 39, Tashkent 100000, Uzbekistan} 

\author{Kimet Jusufi}
\email{kimet.jusufi@unite.edu.mk}
\affiliation{Physics Department, State University of Tetovo, Ilinden Street nn, 1200, Tetovo, North Macedonia}
\author{Mubasher Jamil}
\email{mjamil@sns.nust.edu.pk (corresponding author)}
\affiliation{School of Natural Sciences, National University of Sciences and Technology, Islamabad, 44000, Pakistan}

\author{Ahmadjon Abdujabbarov}
\email{ahmadjon@astrin.uz}
\affiliation{University of Tashkent for Applied Sciences, Str. Gavhar 1, Tashkent 100149, Uzbekistan}
\affiliation{Tashkent State Technical University, Tashkent 100095, Uzbekistan}

\begin{abstract}
We explore the influence of the axion-plasmon on the optical properties of the charged black hole and show how the axion-plasmon coupling modifies the motion of photons around the charged black hole. We then explore in details observational effects such as the  black hole shadow, the gravitational deflection angle, Einstein rings and shadow images obtained by radially infalling gas on a black hole within a plasma medium. An important finding is that the intensity of the electromagnetic radiation increases with the increase of charge and the size of the black hole shadow decreases with increase of the electric charge for a fixed axion-plasmon coupling values when observed from sufficiently large distance.  Overall, for a constant value of charge the optical appearance of the black hole shadow depends on the surrounding plasma model and the largest  shadow radius if found for the case of no plasma, while the smallest shadow radius is found for the case of homogeneous plasma.
\end{abstract}
\maketitle
\section{Introduction}

Black holes (BH) are an astonishing prediction of Einstein's theory of general relativity due to their immense implications. They are not only central character of most of the high energy activities in the astronomical domain such as active-galactic nuclei, jet emissions, accretion disks, quasi-periodic oscillations and gravitational wave production to name a few, but also have fundamental role to explain the nature of space and time \cite{Bambi:2017khi,Smolin:2000af}. It is expected that several laws of any theory of quantum gravity would be formulated using black hole physics. In this direction, the well-known and the earliest contributions were made by Bekenstein, Penrose, Cartar and Hawking. In particular, Hawking proposed the phenomenon of black hole evaporation suggesting that BHs are thermal objects possessing quantum properties. In the current-age research, black holes provide main theoretical framework for addressing the information loss paradox, holography and entanglement phenomenon.

The central feature of a BH is an event horizon which is by definition unobservable since no light signal can escape the boundary of event horizon to an external observer. To actually probe the spacetime geometry nearest the event horizon, light can traverse a BH in the orbits called photon-rings which constitute the photon sphere \cite{Virbhadra:1999nm,Younas:2015sva,AbhishekChowdhuri:2023ekr}, whereas the dark region inside the photon sphere is termed as the black hole shadow, see \cite{Perlick:2021aok} for a review. The geometry of dark region or the BH shadow effectively (such as its diameter and circularity) depends on the black hole parameters such as mass, charge, spin notably besides the observers viewing angle and distance between them. In the past few years, the Event Horizon Telescope has observed very closely the shadows of a central supermassive BH at the center of M87 galaxy \cite{EventHorizonTelescope:2019dse}, and than later the supermassive BH at the heart of our Milky Way galaxy Sgr A* \cite{EventHorizonTelescope:2022wkp}. In the former case, the astronomers have found the enormous magnetic field structure, plasma accretion and determined the accretion rate of M87 central black hole \cite{EventHorizonTelescope:2021srq}. Hence the detection of a dark shadow is clearly a confirmation of the presence of supermassive black hole at the galactic center and another validation of the theory of general relativity. In literature, several modified gravity and quantum gravity-motivated BHs have been studied to investigate their shadow and constraints on their free parameters have been imposed using EHT data \cite{Atamurotov:2022iwj,Jusufi:2022loj,Jusufi:2021lei,Nampalliwar:2021tyz,Ghasemi-Nodehi:2020oiz,Jusufi:2020odz,Liu:2020ola,Zhu:2019ura}.

Motivated by above mentioned recent discoveries, we study the effects of plasma and magnetic field on the shadow of an electrically charged BH using the Reissner-Nordstr\"om (RN) BH. In particular, BHs can acquire electric charge through a combination of BH spin and the magnetic field. In literature, numerous attempts have been made to constrain the electric charge parameter using the EHT data for Sgr A* and M87* \cite{Ghosh:2022kit,Zajacek:2018ycb,Zakharov:2021gbg,Zajacek:2018vsj}, where it is concluded that the upper limit on the charge of Sgr A* has the following observational upper bound $3\times10^8$C. Similarly, the BH charge has been constrained via measurements of gravitational waves data obtained form the BH-BH or binary merger events \cite{Wang:2020ori}. All these studies confirm that the electric charge on the BH cannot be ruled out. However, the same studies reveal that Kerr and Kerr-Newman black holes are not easily distinguished in a substantial region of the EHT-constrained parameter space. Here we extend our previous model about axion-plasma effects on the Schwarzschild BH to include the electric charge \cite{Atamurotov:2021cgh}. This model help us investigate the effects of both plasma and magnetic field on the RN BH shadow.

The plan of the paper is as follows. In Sec.~\ref{Sec:geodesics} we consider the motion of a photon around a BH in the presence of a magnetized plasma, and an axion-plasmon effect on the BH's shadow is studied in more detail in Sec.~\ref{Sec:shadow}. In Sec.~\ref{Sec:lensing} we also consider the optical properties around a BH, which is a gravitational lensing in the presence of magnetized plasma. In Sec.~\ref{sec:massive}, deflection of massive particles and their correspondence with the light rays are considered with more details in presence of magnetized plasma. The Einstein rings in the weak field limit are studied in Sec.~\ref{sec:ring}. In the Sec. \ref{infalling}, we study the effects of gas accretion in the presence of plasma on a BH and determine associated impact on the BH shadow. Finally, we discuss our results in Sec.~\ref{Sec:conclusion}.   
Throughout the paper, we use a system of geometric units in which $G = 1 = c$. Greek indices run from $0$ to $3$.

\section{Photon motion around the BH in the presence of axion-plasmon}
\label{Sec:geodesics}
We investigate a generalized electromagnetic theory that takes the axion-photon interaction into account \cite{Mendonca:2019eke,PhysRevLett.58.1799}.
We have an additional term, due to the interaction between axion and photon proportional to the magnetic field.   The Einstein-Maxwell action is given by 
\begin{equation}
    S=\int \left(\frac{R}{16 \pi}- \frac{1}{4}F_{\mu\nu}F^{\mu\nu}\right) \sqrt{-g}\, d^4x+S_{\rm matter},
\end{equation}
where 
\begin{eqnarray}
    S_{\rm matter}=\int \mathcal{L}_{\rm matter}\sqrt{-g}\, d^4x,
\end{eqnarray}
with 
\begin{eqnarray}
    \mathcal{L}_{\rm matter}=\mathcal{L}_\varphi-A_\mu J_e^\mu+\mathcal{L}_{\text{int}}.
\end{eqnarray}
In this work we assume that the surrounding plasma has no strong effect on the spacetime geometry, however the presence of plasma has an influence on the photons by modifying the energy of the photons. In the above Lagrangian, $J_e^\mu$, and $F_{\mu\nu}$,  represent four vector current of electrons and electromagnetic tensor respectively while $\mathcal{L}_\varphi=\nabla_\mu\varphi^*\nabla^\mu\varphi-m_\varphi^2|\varphi|^2$, defines the axion related Lagrangian density, further $\mathcal{L}_{\text{int}}=-(g/4)\varepsilon^{\mu\nu\alpha\beta}F_{\alpha\beta}F_{\mu\nu}$, is the photon-axion connection term, where the appropriate axion-photon coupling is indicated by $g$.

The static spherically symmetric spacetime metric for the Reissner–Nordström BH is provided as
\begin{equation}\label{metric}
ds^2=-f(r)dt^2+\frac{1}{f(r)}dr^2+r^2(d\theta^2+\sin^2{\theta}d\phi^2),
\end{equation}
where $f(r)=1-(2M/r)+(Q^2/r^2)$. Further, $M$ and $Q$ are the candidates for the mass and charge of the BH, respectively.

The Hamiltonian for a photon orbiting about a BH in the presence of axion-plasmon medium is defined as  \cite{Synge:1960b}
\begin{equation}
\mathcal H(x^\alpha, p_\alpha)=\frac{1}{2}\left[ g^{\alpha \beta} p_\alpha p_\beta - (n^2-1)( p_\beta u^\beta )^2 \right],
\label{generalHamiltonian}
\end{equation}
where $x^\alpha$ define the spacetime coordinates, $u^\beta$ and $p_\alpha$ are the four-velocity and four-momentum of the photon respectively. In the above expression $n$ represents the refractive index of the plasma medium ($n=\omega/k$, where $k$ is the wave number). The refractive index in the background of an axion-plasmon contribution is expressed as follows~\cite{Mendonca:2019eke}
\begin{eqnarray}
n^2&=&1- \frac{\omega_{\text{p}}^2}{\omega^2}-\frac{f_0}{\gamma_{0}}\frac{\omega_{\text{p}}^2}{(\omega-k u_0)^2}-\frac{\Omega^4}{\omega^2(\omega^2-\omega_{\varphi}^2)}\nonumber \\
&&-\frac{f_0}{\gamma_{0}}\frac{\Omega^4}{(\omega-k u_0)^2(\omega^2-\omega_{\varphi}^2)},
\label{eq:n1}
\end{eqnarray}
in terms of the plasma frequency $\omega^2_{p}(x^\alpha)=4 \pi e^2 N(x^\alpha)/m_e$ ($e$ and $m_e$ are the electron charge and mass respectively whereas $N$ is the number density of the electrons), the photon frequency $\omega(x^\alpha)$ is defined by $\omega^2=( p_\beta u^\beta )^2$, the axion frequency $\omega_{\varphi}^2$%
, the axion-plasmon coupling parameter $\Omega=(gB_{0}\omega_{p})^{1/2}$ with $B_0$ being the homogeneous magnetic field in the $z$-direction. Because the magnetic field is axial symmetric and the spacetime geometry is spherically symmetric, we presume that the strong spacetime curvature play the dominant role in particle dynamics in comparison with magnetic field. The parameter $f_0$ is the fraction of the electrons in the beam propagating inside the plasma with velocity $u_0$ and $\gamma_0$ is the corresponding Lorentz factor. Because the role of the electron beam scenario near the BH is less clear, we set $f_0=0$ for simplicity and rewrite~(\ref{eq:n1}) as \cite{Atamurotov:2021cgh}
\begin{eqnarray}
n^2(r)&=&1- \frac{\omega_{\text{p}}^2(r)}{\omega(r)^2}-\frac{\Omega^4}{\omega(r)^2[\omega(r)^2-\omega_{\varphi}^2]},\nonumber \\
&&=1- \frac{\omega_{\text{p}}^2(r)}{\omega(r)^2}\left(1+\frac{ g^2B^2_0 }{\omega(r)^2-\omega_{\varphi}^2}\right)
,
\label{eq:n2}
\end{eqnarray}
with
\begin{equation}
\omega(r)=\frac{\omega_0}{\sqrt{f(r)}},\qquad  \omega_0=\text{const}.
\end{equation}
Experiments concerning the axion-plasmon conversion impose the following constraint upon frequency scales $\omega_{\text{p}}^2\gg \Omega^2$ or $\omega_{\text{p}}\gg gB_0,$ where $g$ denotes the axion-photon coupling \cite{Mendonca:2019eke}. The lapse function is such that $f(r) \to 1$ as $r \to \infty$ and $\omega(\infty)=\omega_0=-p_t,$ which represents energy of the photon at spatial infinity \cite{Perlick2015}. Besides, the plasma frequency must be sufficiently small than the photon frequency $(\omega_{\text{p}}^2\ll \omega^2)$ which allows the BH shadow to be differentiated from the vacuum case. The Hamiltonian for the light rays in the axion-plasmon medium has the form 
\begin{equation}
\mathcal{H}=\frac{1}{2}\Big[g^{\alpha\beta}p_{\alpha}p_{\beta}+\omega^2_{\text{p}}\Big(1+\frac{ g^2B^2_0 }{\omega^2_0-\omega_{\varphi}^2}\Big)\Big]. \label{eq:hamiltonnon}
\end{equation}
 The components of the four velocity for the photons in the equatorial plane $(\theta=\pi/2,~p_\theta=0)$ are given by
\begin{eqnarray} 
\dot t\equiv\frac{dt}{d\lambda}&=& \frac{ {-p_t}}{f(r)}  , \label{eq:t} \\
\dot r\equiv\frac{dr}{d\lambda}&=&p_rf(r) , \label{eq:r} \\
\dot\phi\equiv\frac{d \phi}{d\lambda}&=& \frac{p_{\phi}}{r^2}, \label{eq:varphi}
\end{eqnarray}
where we used the relationship, $\dot x^\alpha=\partial \mathcal{H}/\partial p_\alpha$. From Eqs. (\ref{eq:r}) and (\ref{eq:varphi}), we obtain a governing equation for the phase trajectory of light
\begin{equation}
\frac{dr}{d\phi}=\frac{g^{rr}p_r}{g^{\phi\phi}p_{\phi}}.    \label{trajectory}
\end{equation}
Using the constraint $\mathcal H=0$, we can rewrite the above equation as~\cite{Perlick2015}
\begin{equation}
 \frac{dr}{d\phi}=\sqrt{\frac{g^{rr}}{g^{\phi\phi}}}\sqrt{h^2(r)\frac{\omega^2_0}{p_\phi^2}-1},
\end{equation}
where we defined
\begin{equation}
    h^2(r)\equiv-\frac{g^{tt}}{g^{\phi\phi}}-\frac{\omega^2_p}{g^{\phi\phi}\omega^2_0}\left(1+\frac{ g^2B^2_0 }{\omega^2_0-\omega_{\varphi}^2}\right).
\end{equation}
We now introduce the dimensionless parameters
\begin{equation}\label{dim}
\tilde \omega_{\varphi}^2=\frac{\omega_{\varphi}^2}{\omega^2_0} \quad\text{and}\quad \tilde B^2=\frac{g^2B^2_0}{\omega^2_0},
\end{equation}
which yield
\begin{equation}
h^2(r)=r^2\Big[\frac{r^2}{r^2-2 M r+Q^2}-\frac{\omega^2_{\text{p}}(r)}{\omega^2_0}\Big(1+\frac{ \tilde B^2 }{1-\tilde \omega_{\varphi}^2}\Big)\Big]. \label{eq:hrnew}
\end{equation}
The radius of a circular orbit of light, particularly the one which forms the photon sphere of radius $r_{\text{ph}}$, is determined by solving the following equation ~\cite{Perlick2015}
\begin{equation}
\frac{d(h^2(r))}{dr}\bigg|_{r=r_{\text{ph}}}=0. \label{eq:con}    
\end{equation}
By substituting Eq.~(\ref{eq:hrnew}) in (\ref{eq:con}) one can write the algebraic equation for $r_{\text{ph}}$ in the presence of plasma medium as
\begin{eqnarray}\nonumber\label{eq:orbits}
\Big[\frac{\omega^2_{\text{p}}(r_{\text{ph}})}{\omega^2_0}+&\frac{r_{\text{ph}}\omega'_{\text{p}}(r_{\text{ph}}) \omega_{\text{p}}(r_{\text{ph}})}{\omega^2_0}\Big]\Big(1+\frac{ \tilde B^2_0 }{1-\tilde \omega_{\varphi}^2}\Big)\\
&=\frac{r_{\text{ph}}^4-3 r_{\text{ph}}^3 M+2Q^2 r_{\text{ph}}^2}{(r_{\text{ph}}^2-2 M r_{\text{ph}}+Q^2)^2},
\end{eqnarray}
where prime denotes the derivative with respect to radial coordinate $r$. Although the roots of Eq. (\ref{eq:orbits}) cannot be obtained analytically for most choices of $\omega_{\text {p}}(r)$, we can plot graphs for the specific options of $\omega_{\text {p}}(r)$ numerically. 

\begin{figure}
 \begin{center}
   \includegraphics[scale=0.45]{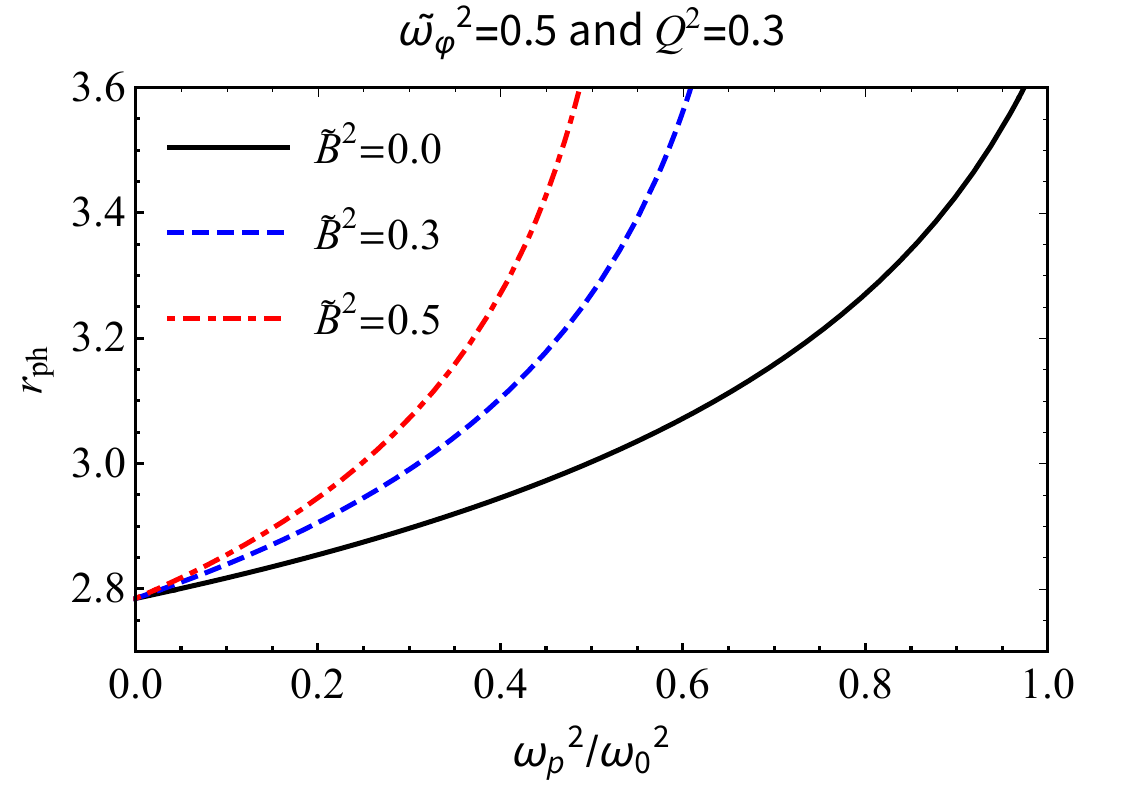}
   \includegraphics[scale=0.45]{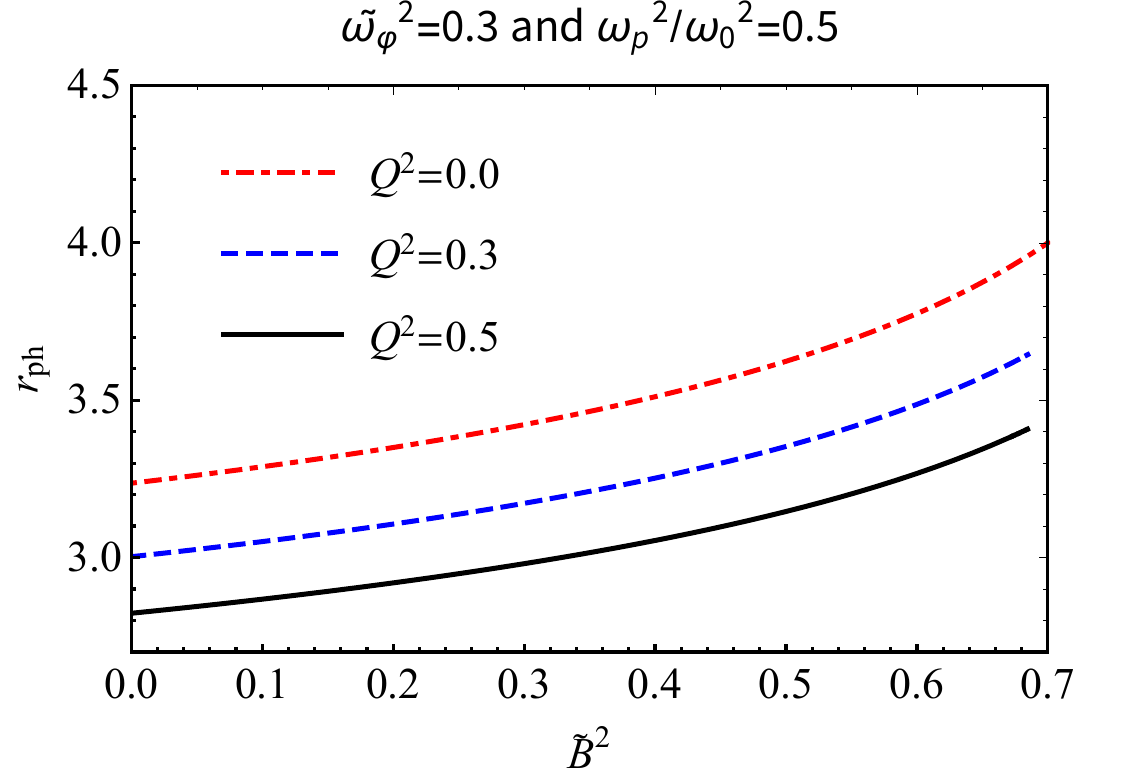}
    \includegraphics[scale=0.45]{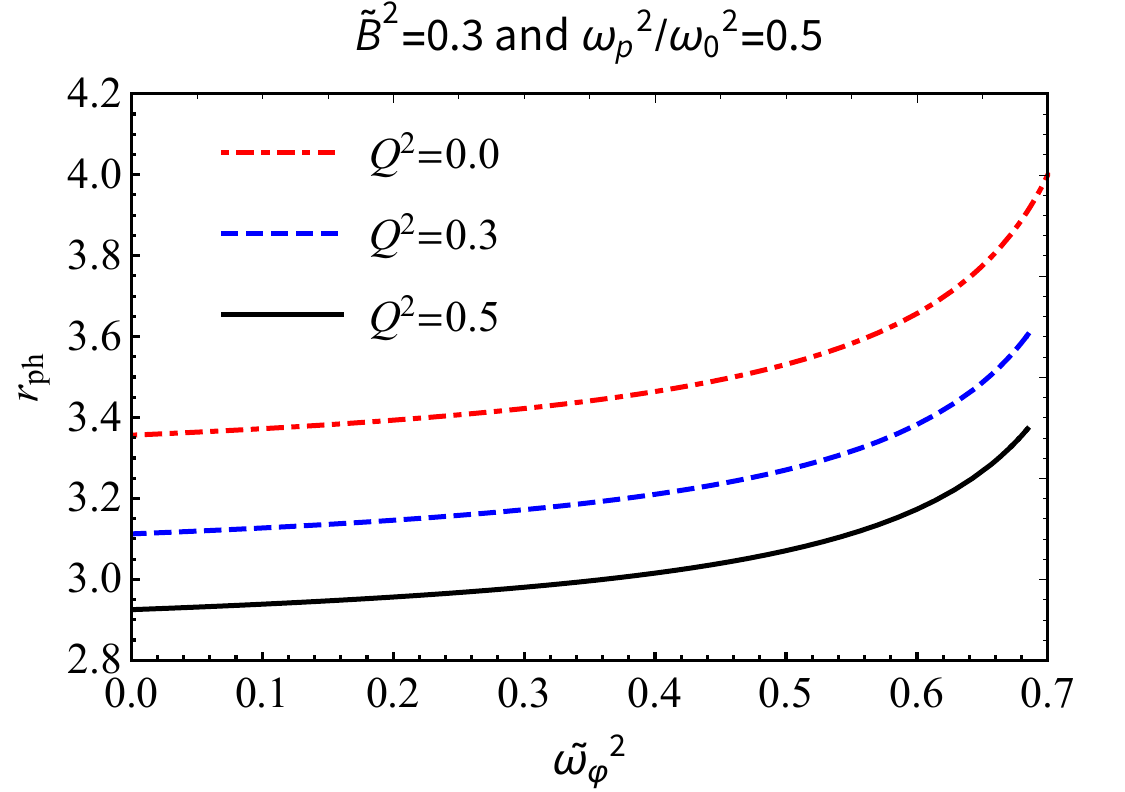}
    \includegraphics[scale=0.45]{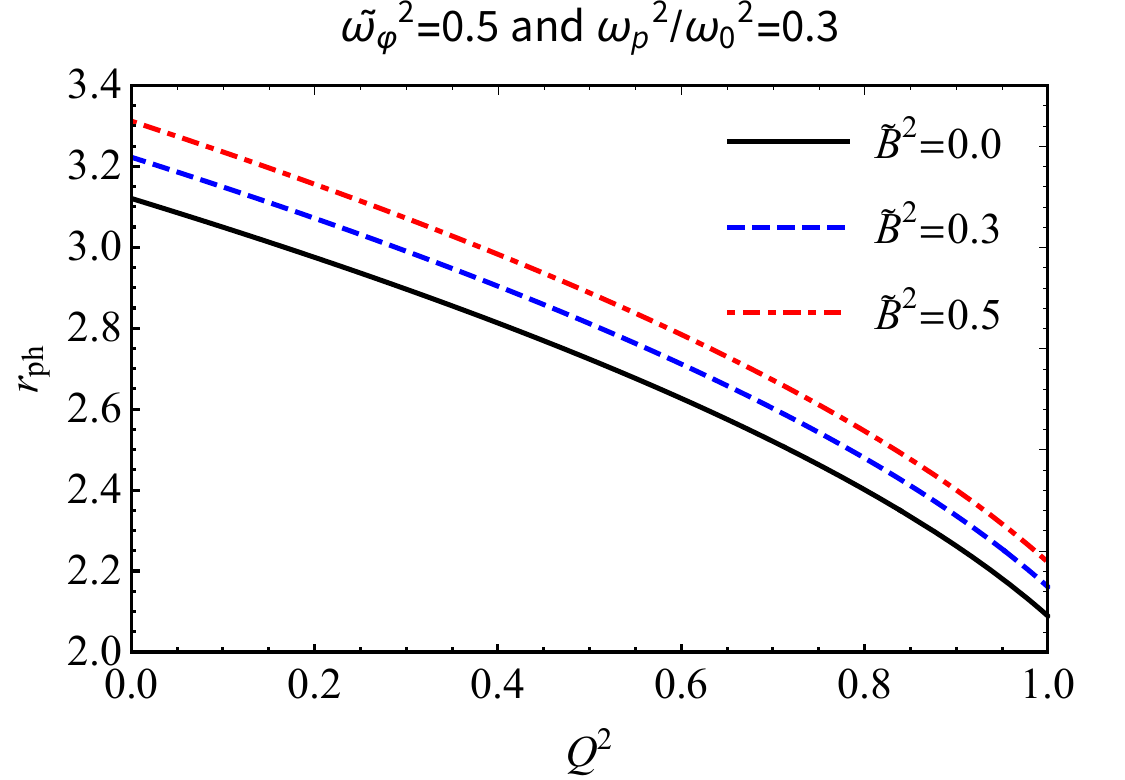}
  \end{center}
\caption{Radius of the photon sphere for the homogeneous plasma with axion field.}\label{plot:photonradiusuni}
\end{figure}

\subsection{Homogeneous plasma with $\omega _p^2(r)= \text{const.}$} \label{2.A}

In the special case of a homogeneous plasma with axions or fixed plasma frequency i.e. $\omega _p^2= \text{const.}$, the exact form of the photon sphere cannot be obtained, because of this we plot the graph by sitting some values for $\tilde{\omega}_{\phi}^2,\omega_p^2/\omega_0^2, Q^2$ and $ \tilde{B}^2$. For simplicity, we set $M=1$, this operation effectively corresponds to rescaling the Boyer- Lindquist radial coordinate and electric charge as $\tilde{r}= r/M$ and $\tilde{Q} = Q/M$ respectively. Throughout this work we will drop the tildes.  Plots of $r_{\text{p}}$ are depicted in Fig.~\ref{plot:photonradiusuni} versus the plasma frequency, the magnetic field, the axion frequency and the charge of the black hole seperately. The figures suggest that while the first three physical factors contribute to increase the size of the photon sphere, the charge effects conversely.


\subsection{Inhomogeneous plasma with $\omega^2_{p}(r)=z_0/r^q$}
\begin{figure}
 \begin{center}
   \includegraphics[scale=0.46]{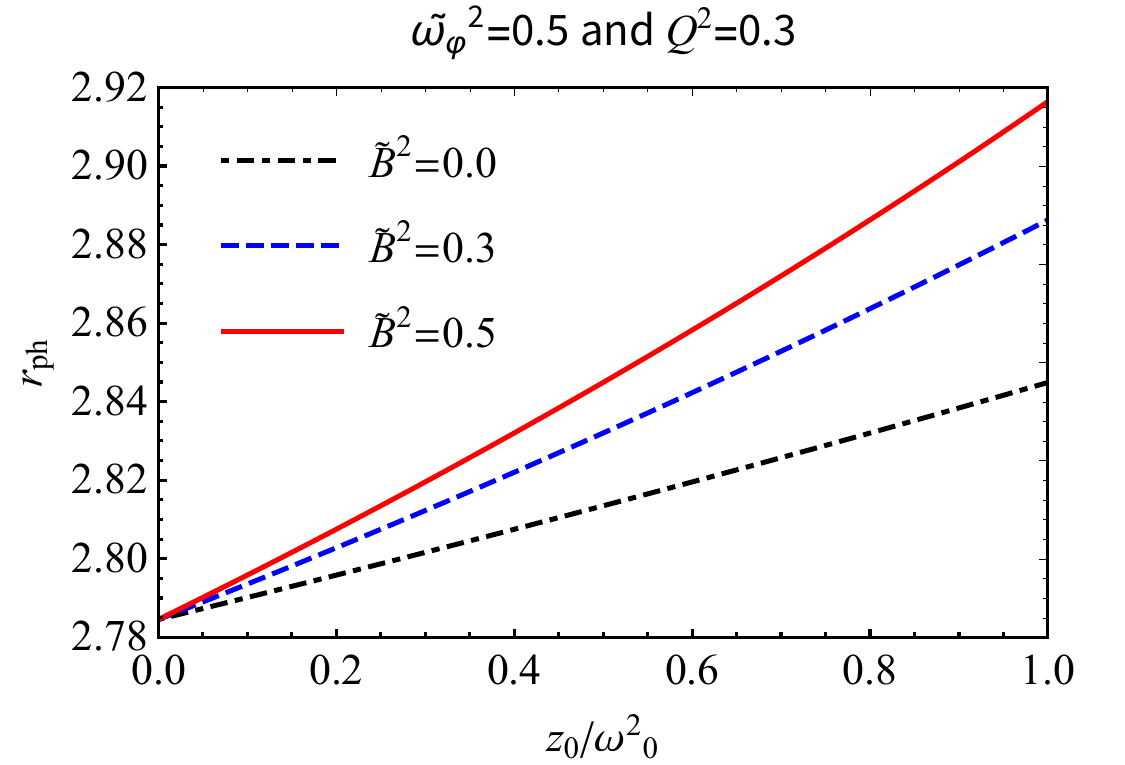}
   \includegraphics[scale=0.46]{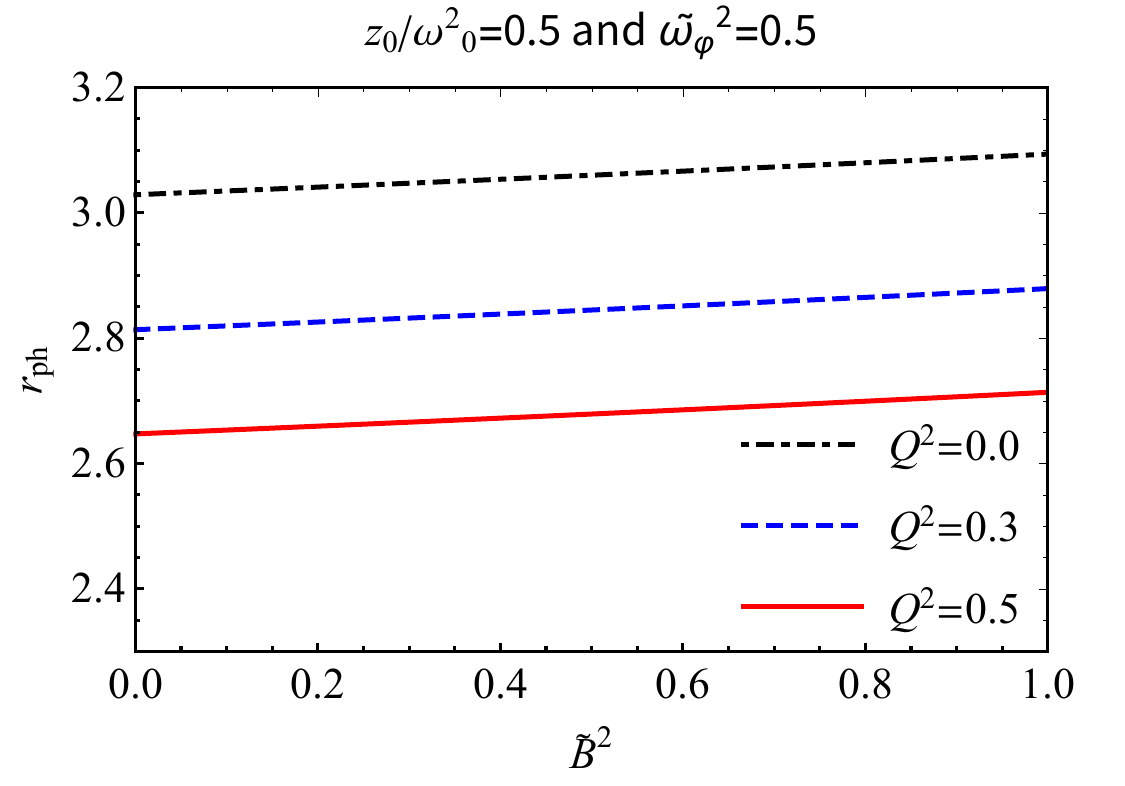}
    \includegraphics[scale=0.46]{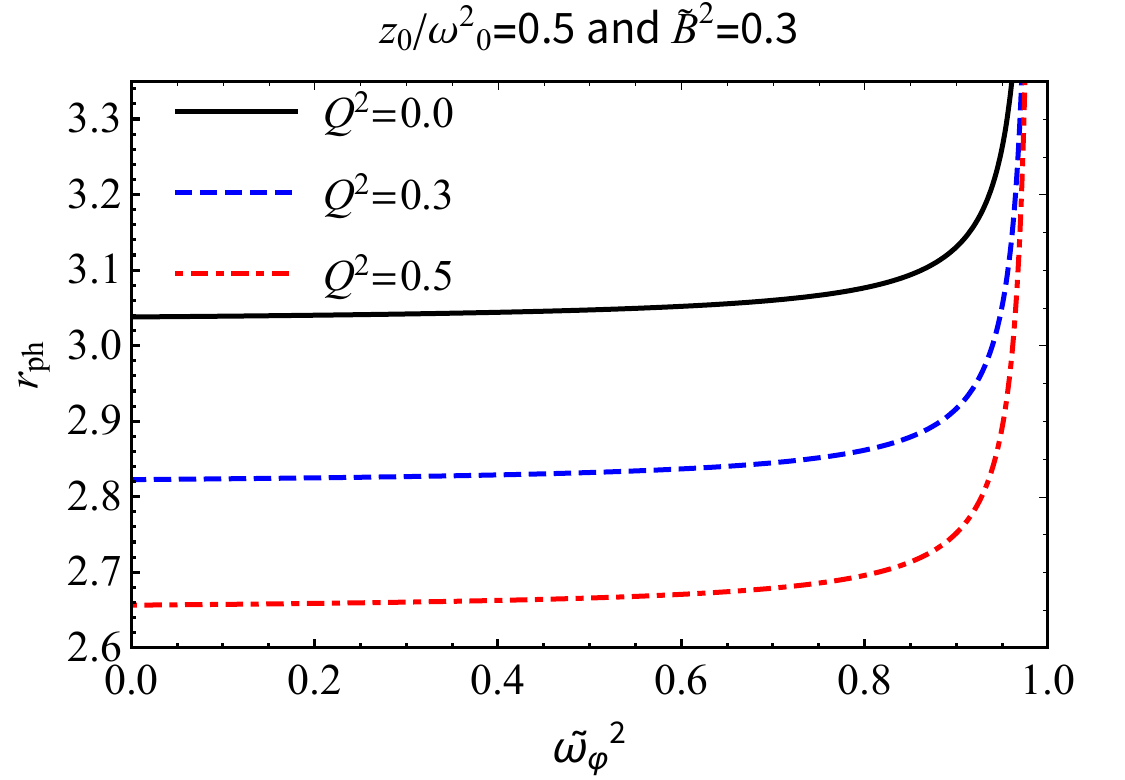}
    \includegraphics[scale=0.46]{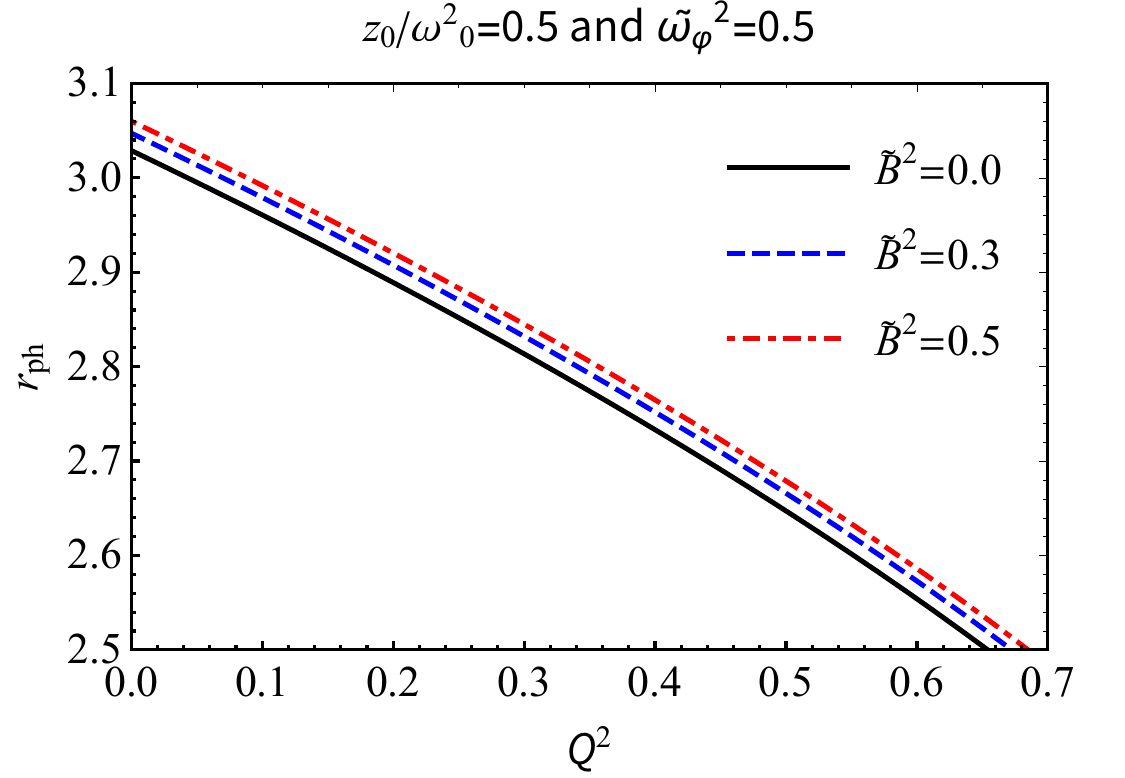}
  \end{center}
\caption{Radius of photon sphere for the inhomogeneous power-law plasma with axion field.}\label{plot:photonradiusnonuni}
\end{figure}

Now we explore photon spheres in the presence of an inhomogeneous plasma with axion, where the plasma frequency is required to satisfy a simple power-law of the form \cite{Rog:2015a,Er2017aa}
\begin{equation}\label{eq:omegaplasma}
\omega^2_{p}(r)=\frac{z_0}{r^q},
\end{equation}
where $z_0$ and $q$ are free parameters.
To analyze the main features of the power-law model we restrict ourselves to the case  $q=1$ and $z_0$ as a constant \cite{Rog:2015a}.
Again we have plotted graphs numerically and chose the solution of [\ref{eq:orbits}] when there is no plasma contribution ($z_0=0$), then the root reduces to $r_{\text{ph}}=\frac{1}{2}\big(3M+\sqrt{9 M^2-8 Q^2}\big)$, which is the photon sphere for a non-rotating charged BH. In Fig.~\ref{plot:photonradiusnonuni} the radius of photon sphere versus different free parameters are given. The effects of parameters of axion-plasmon model on the size of photon sphere are evidently manifested. The radius of photon radius decreases with the increase of the electric charge.

\section{BH shadow in an axion-plasmon medium }
\label{Sec:shadow}
\begin{figure}
 \begin{center}
   \includegraphics[scale=0.455]{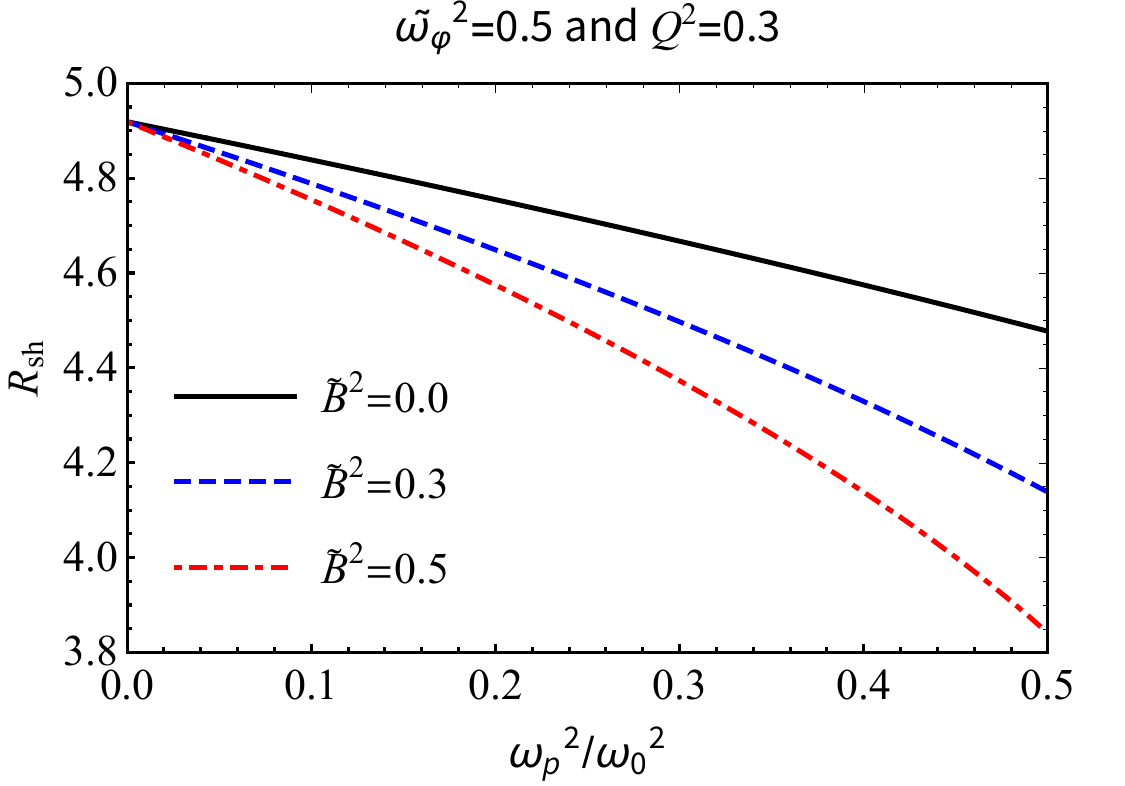}
   \includegraphics[scale=0.455]{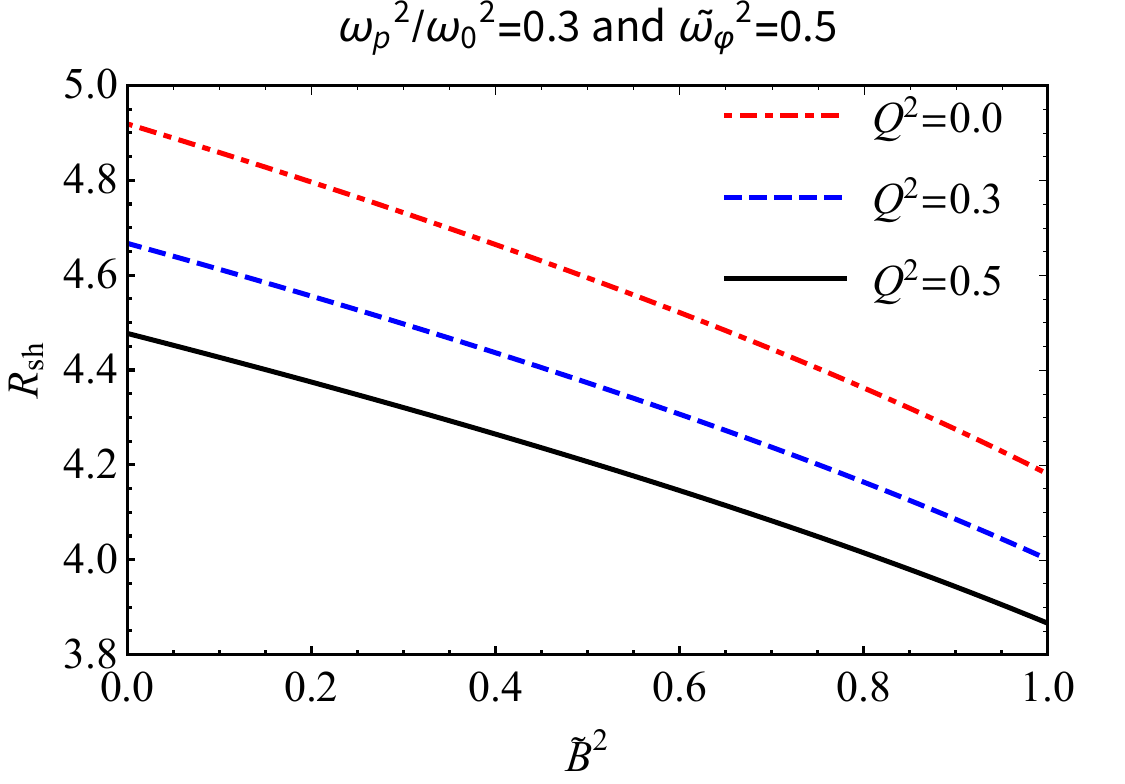}
    \includegraphics[scale=0.455]{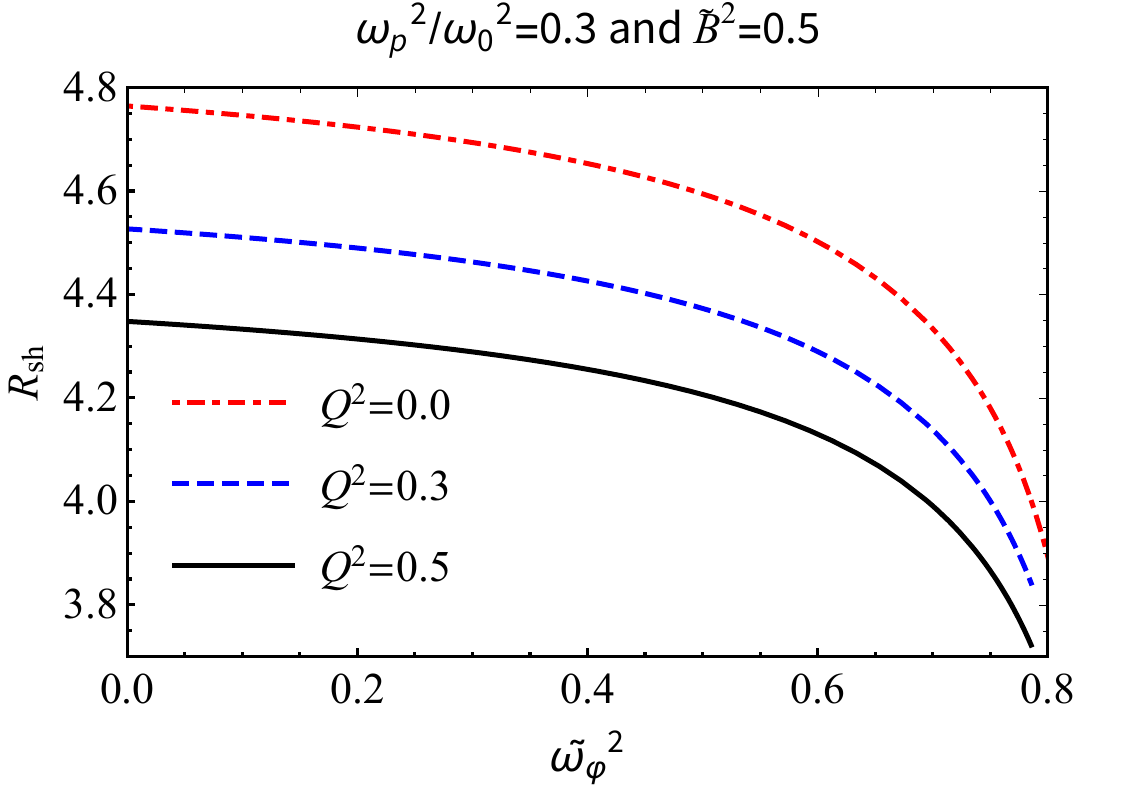}
    \includegraphics[scale=0.455]{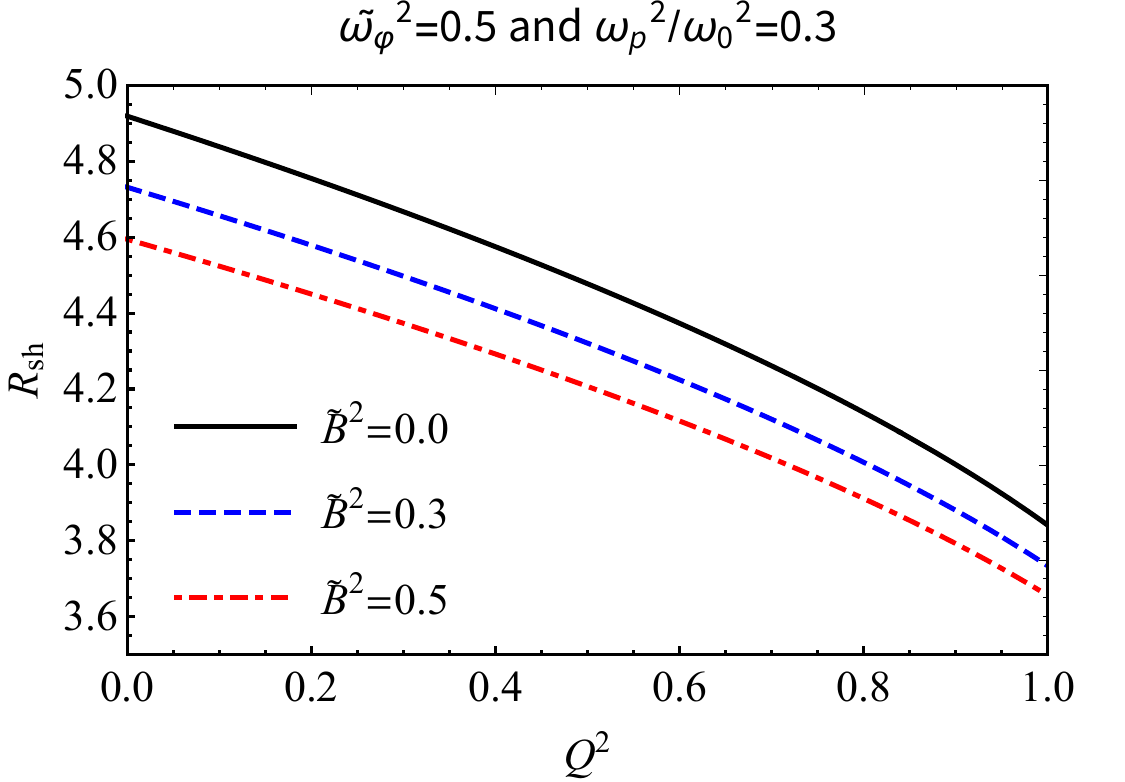}
  \end{center}
\caption{Shadow's radius of the BH for the homogeneous constant-frequency plasma with axion field}\label{plot:shadowuni}
\end{figure}

\begin{figure}
 \begin{center}
   \includegraphics[scale=0.45]{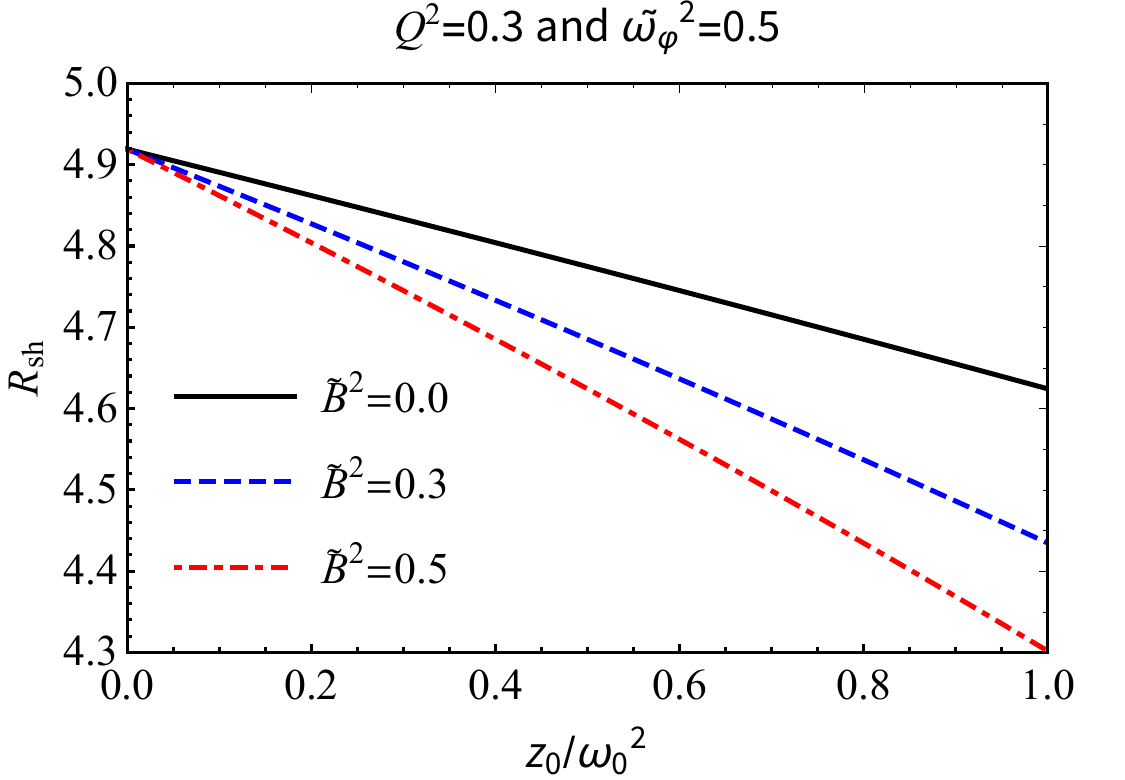}
   \includegraphics[scale=0.45]{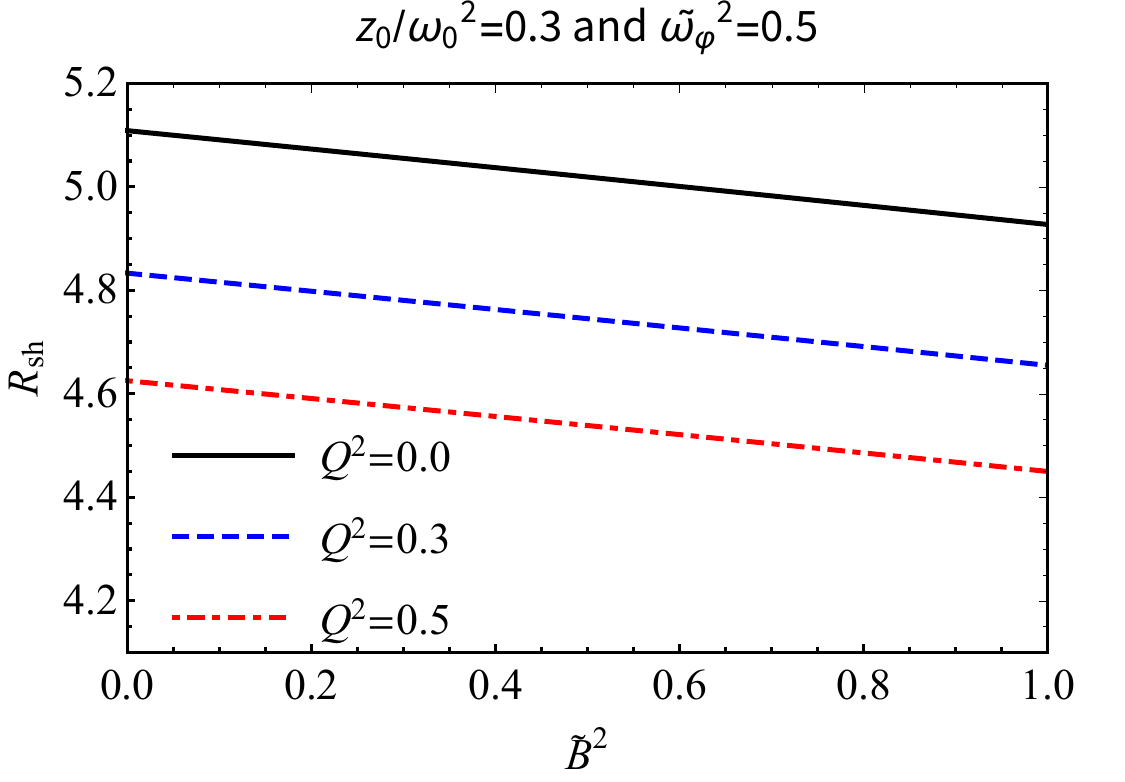}
    \includegraphics[scale=0.45]{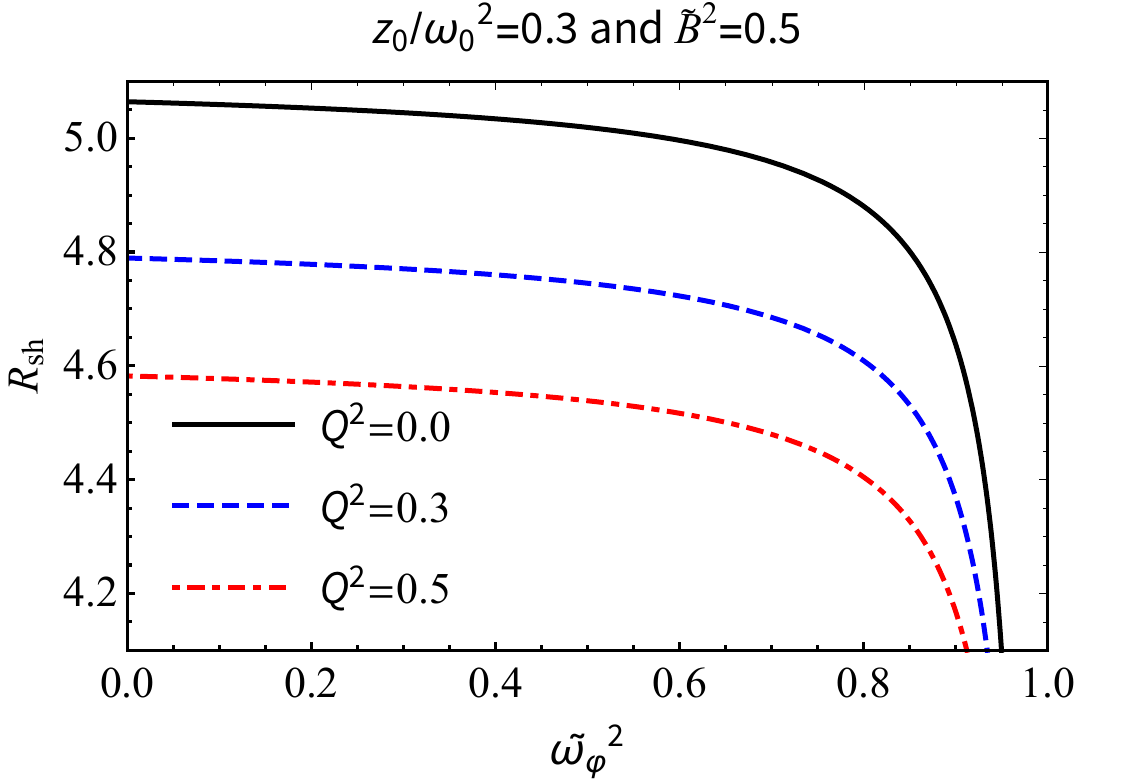}
    \includegraphics[scale=0.45]{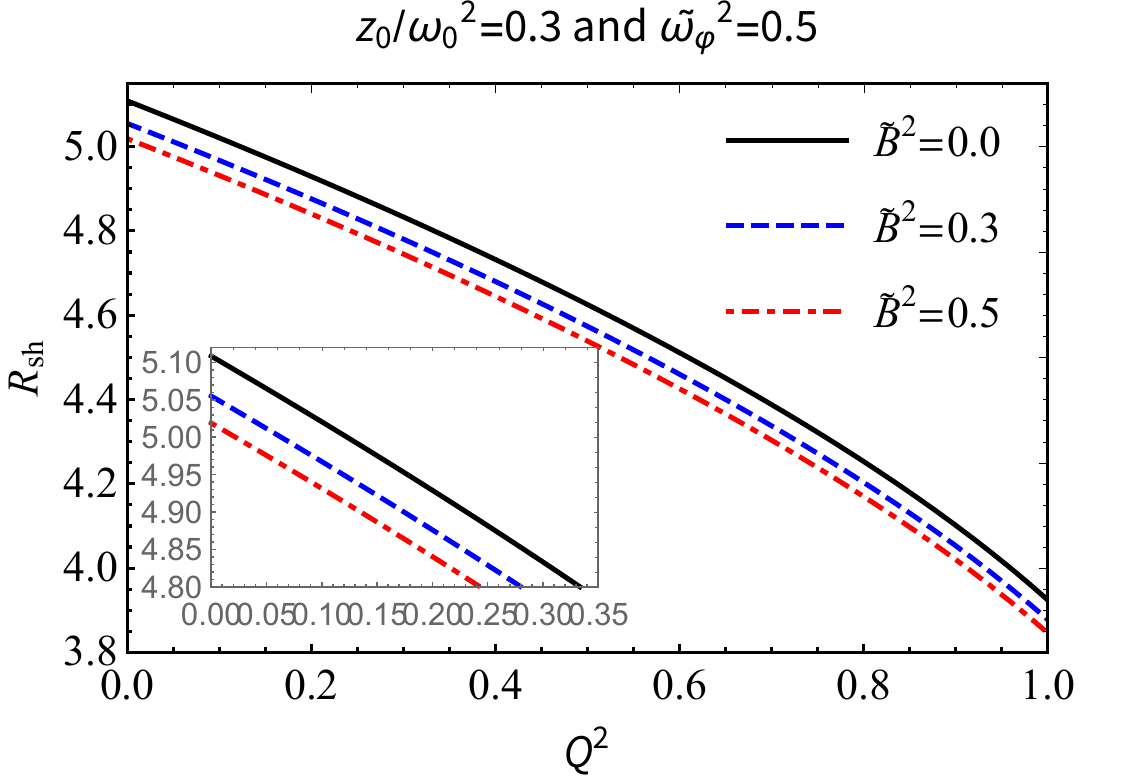}
  \end{center}
\caption{Radius of the BH shadow for the inhomogeneous plasma with axion field and power-law plasma frequency.}\label{plot:shadownonuni}
\end{figure}

In this section we study the radius of the shadow of the Reissner–Nordström metric in the presence of a magnetized plasma. The angular radius $\alpha_{\text{sh}}$ of the BH shadow is defined by a geometric approach which results in~\cite{Synge66,Perlick2015}
\begin{eqnarray}
\label{eq:shadow nonrotating1}
\sin^2 \alpha_{\text{sh}}&=&\frac{h^2(r_{\text{ph}})}{h^2(r_{\text{o}})},\\&=&\frac{r_{\text{ph}}^2\left[\frac{r_{\text{ph}}^2}{r_{\text{ph}}^2-2 M r_{\text{ph}}+Q^2}-\frac{\omega^2_p(r_{\text{ph}})}{\omega^2_0}\left(1+\frac{\tilde{B}^2}{1-\tilde{\omega }_{\varphi }^2}\right)\right]}{r_{\text{o}}^2\left[\frac{r_{\text{o}}^2}{r_{\text{o}}^2-2 M r_{\text{o}}+Q^2}-\frac{\omega^2_p(r_{\text{o}})}{\omega^2_0}\left(1+\frac{\tilde{B}^2}{1-\tilde{\omega }_{\varphi }^2}\right)\right]},\nonumber 
\end{eqnarray}
where $r_{\text{o}}$ and $r_{\text{ph}}$ represent the locations of the observer and the photon sphere respectively. If the observer is located at a sufficiently large distance from the BH then one can approximate radius of BH shadow by using Eq.~(\ref{eq:shadow nonrotating1}) as~\cite{Perlick2015}
\begin{eqnarray}
R_{\text{sh}}&\simeq& r_{\text{o}} \sin \alpha_{\text{sh}},\\
 &=&\sqrt{r_{\text{ph}}^2\bigg[\frac{r_{\text{ph}}^2}{r_{\text{ph}}^2-2 M r_{\text{ph}}+Q^2}-\frac{\omega^2_p(r_{\text{ph}})}{\omega^2_0}\bigg(1+\frac{\tilde{B}^2}{1-\tilde{\omega }_{\varphi }^2}\bigg)\bigg]},  \nonumber
\end{eqnarray}
where we have used the fact that $h(r)\to r$, which follows from Eq. (\ref{eq:hrnew}), at spatial infinity for both models of plasma along with a constant magnetic field.
In the case of vacuum $\omega_{\text{p}}(r)\equiv0$, and considering $Q=0$ we recover the radius of Schwarzschild BH shadow $R_{\text{sh}}=3\sqrt{3} M$ when $r_{\text{ph}}=3M$.
The radius of BH shadow is depicted for different parameters in Fig.~\ref{plot:shadowuni} for a homogeneous plasma with fixed plasma frequency and Fig.~\ref{plot:shadownonuni} shows the case for a power-law model of plasma frequency $\omega^2_{p}(r)=z_0/r$. We observe that the size of shadow radius decreases by increasing the magnetic field strength or the axion-plasmon frequency. Thus the BH shadow in the presence of axion-plasmon medium would shrink further,  as expected.
It is interesting to note that the effects of a homogeneous plasma on the radius of the BH shadow are more pronounced than the effects of an inhomogeneous plasma, which is abundantly clear from Figs.~\ref{plot:shadowuni} and~Figs. \ref{plot:shadownonuni}.

%
%
%

\section{Gravitational lensing and deflection angle of light in the plasma with axion field}\label{Sec:lensing}

Now we consider the gravitational lensing of light paths in the presence of plasma with axion field. The trajectory is defined by Eq. (\ref{trajectory}) which implies
\begin{eqnarray}
 \frac{d\phi}{dr}= \frac{1}{r^2 f(r)}\frac{p_{\phi}}{p_{r}}. \label{eq:phir}
\end{eqnarray}
Using Eqs. (\ref{eq:hamiltonnon}), (\ref{dim}) and (\ref{eq:phir}), we arrive at
\begin{eqnarray}
 \frac{d\phi}{dr}=\pm \frac{p_{\phi}}{r^2}\frac{1}{\sqrt{p^2_{t}-\frac{f(r)}{r^2}\left[p_\phi^2+r^2 \omega^2_{p}\left(1+\frac{ g^2B^2_0 }{\omega^2_0-\omega_{\varphi}^2}\right)\right]}}, \label{eq:phirlast}
\end{eqnarray}
which results in
\begin{equation}\label{eq:philast2}
\Delta \phi =  2 \int \limits_R^\infty \frac{p_\phi}{r^2 } \, \frac{dr}{\sqrt{p_t^2 - f(r)  \big[\frac{p_\phi^2}{r^2} + \omega_{\text{p}}^2(r)\big(1+\frac{\tilde{B}^2}{1-\tilde{\omega }_{\varphi }^2}\big) \big]} } \, .	
\end{equation}%
In the above equation, it is assumed that the light ray travels from the source at spatial infinity, passing by the BH with the closest approach at $r=R$ and than escaping to later arrive at the observer location at infinity. Due to symmetry of the scenario, we write a factor of $2$ in the above integral.

 The light ray is deflected from a straight line path at the difference
of angle $\pi$ which results in the total deflection angle given by \cite{Weinberg:1972kfs}:
\begin{equation}
\hat{\alpha} = 2 \int \limits_R^\infty \frac{p_\phi}{r^2 } \, \frac{dr}{\sqrt{p_t^2 - f(r)  \big[\frac{p_\phi^2}{r^2} + \omega_{\text{p}}^2(r)\big(1+\frac{\tilde{B}^2}{1-\tilde{\omega }_{\varphi }^2}\big) \big]} }  - \pi \, . \label{eq:defangle}
\end{equation}
Note that $r=R$ is a turning point: $dr/d \lambda = 0$ and $p_r = 0$. The expressions of $p_t^2$ and $p_\phi^2$ at the turning point are, respectively given by 
\begin{equation} \label{eq:boundary}
p_t^2 = f(R)  \Big[ \frac{p_\phi^2}{R^2} + \omega_{\text{p}}^2(R)\Big(1+\frac{\tilde{B}^2}{1-\tilde{\omega }_{\varphi }^2}\Big) \Big] \, ,
\end{equation}
\begin{equation}
p_\phi^2 = R^2 p_t^2 \Big[ \frac{1}{f(R)} - \frac{\omega_{\text{p}}^2(R)}{\omega_0^2} \Big(1+\frac{\tilde{B}^2}{1-\tilde{\omega }_{\varphi }^2}\Big) \Big] \, .
\end{equation}

Using Eqs.~(\ref{eq:hrnew}) and (\ref{eq:phirlast}) we rewrite the equation of trajectory of photons in the Reissner–Nordström spacetime as
\begin{equation} \label{eq:light}
\frac{d \phi}{dr} = \pm \frac{1}{\sqrt{r^2-2M r+Q^2}\sqrt{\frac{h^2(r)}{h^2(R)} - 1 }} \, .
\end{equation}
The deflection angle in the presence of the plasma with axion field assumes the form
\begin{eqnarray} \label{eq:lastdef}
\hat{\alpha} &=& 2 \int \limits_R^\infty \frac{dr}{\sqrt{r^2-2M r+Q^2}\sqrt{\frac{h^2(r)}{h^2(R)} - 1 }} - \pi .
\end{eqnarray}
where 
\begin{equation}
\Big(\frac{h(r)}{h(R)}\Big)^2=\frac{r^2\left[\frac{r^2}{r^2-2 M r+Q^2}-\frac{\omega^2_p(r)}{\omega^2_0}\left(1+\frac{\tilde{B}^2}{1-\tilde{\omega }_{\varphi }^2}\right)\right]}{R^2\left[\frac{R}{R^2-2 M R+Q^2}-\frac{\omega^2_p(R)}{\omega^2_0}\left(1+\frac{\tilde{B}^2}{1-\tilde{\omega }_{\varphi }^2}\right)\right]}.
\end{equation}

\begin{figure}
 \begin{center}
   \includegraphics[scale=0.45]{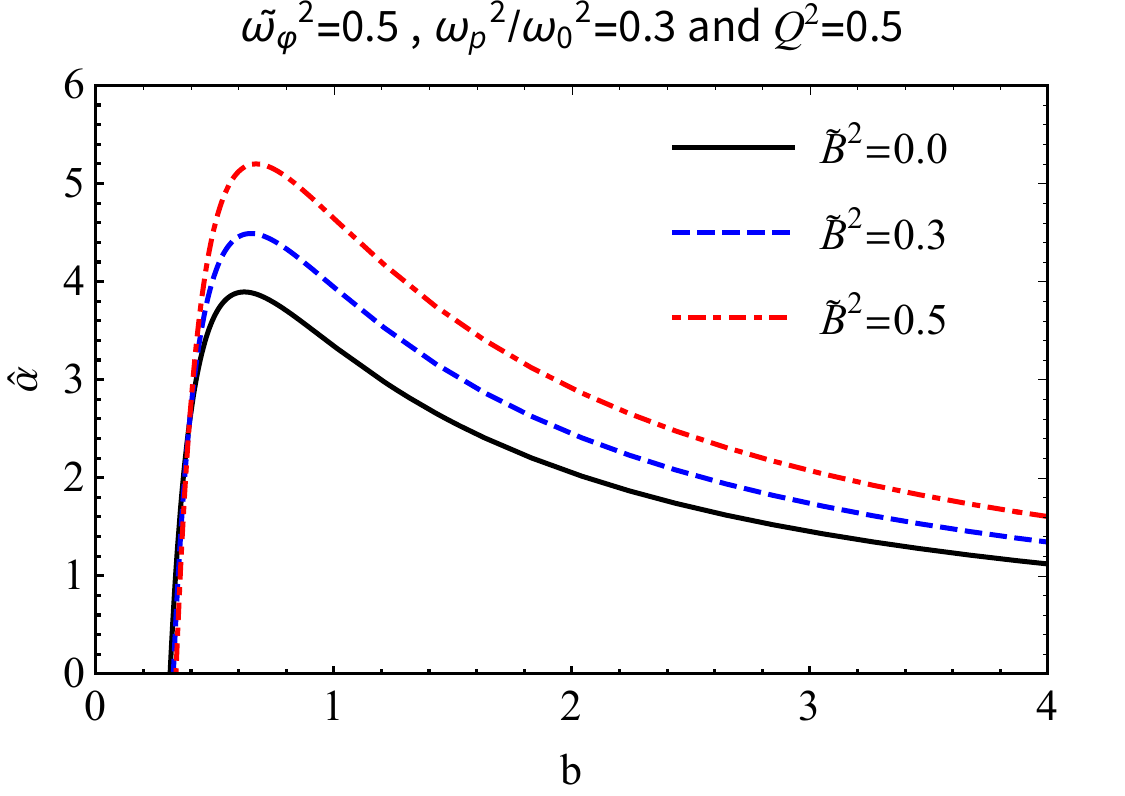}
   \includegraphics[scale=0.45]{avsbQ.pdf}
  \end{center}
\caption{Deflection angle versus the impact parameter $b$ in presence of a plasma axion fluid.}\label{plot:defbuni}
\end{figure}

\begin{figure}
 \begin{center}
   \includegraphics[scale=0.46]{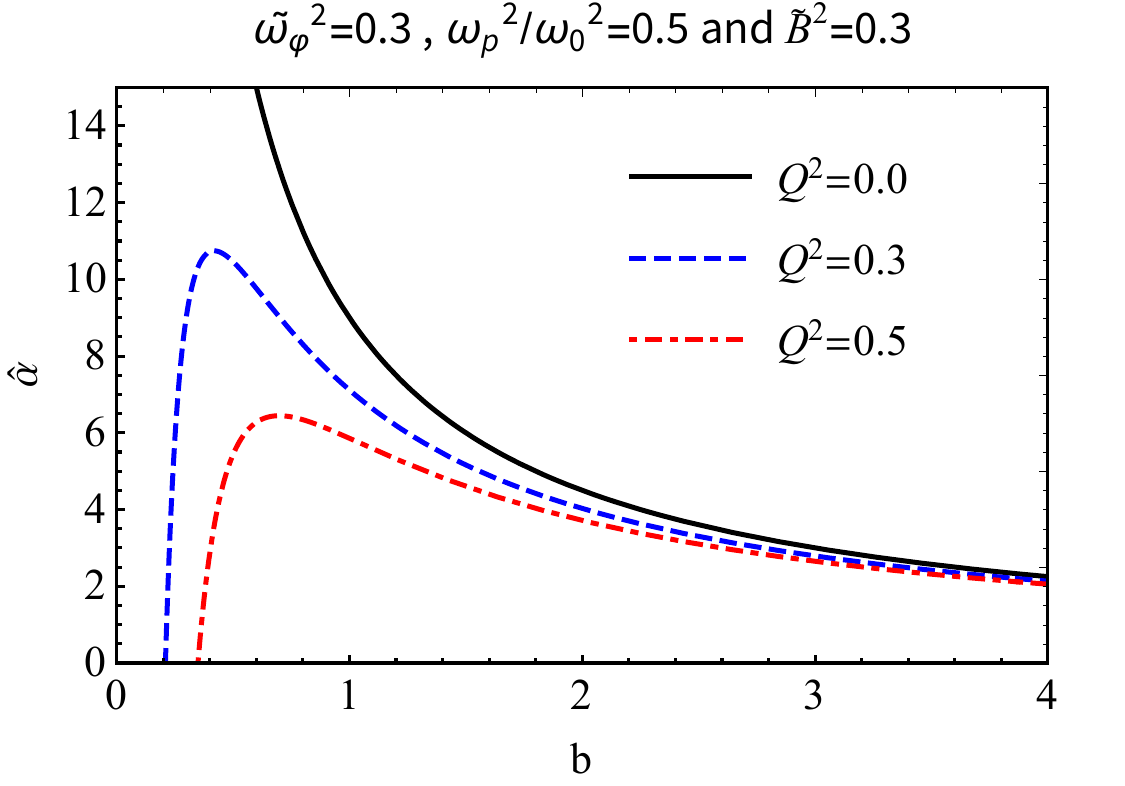}
   \includegraphics[scale=0.46]{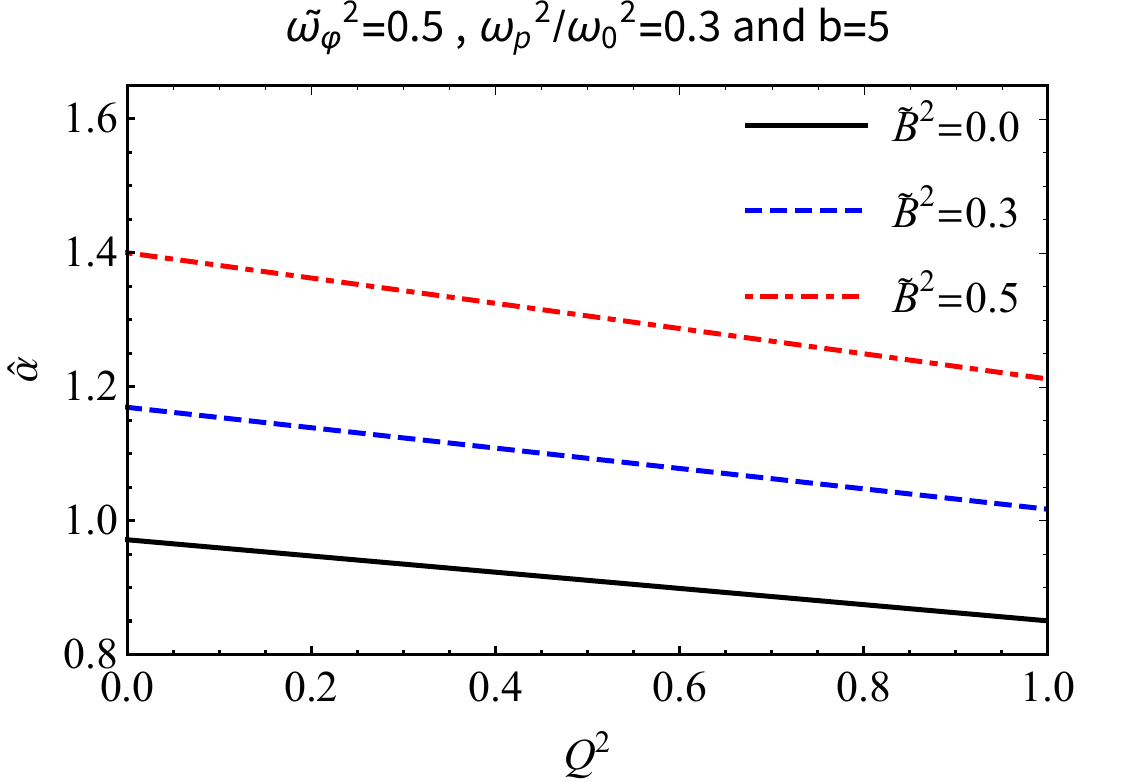}
  \end{center}
\caption{Deflection angle versus the axion fluid parameters.}\label{plot:defuni}
\end{figure}

For $R\gg M$ and a uniform plasma $\omega^2_{p}=\text{const.}$, we get
\begin{eqnarray}\nonumber
\frac{h^2(r)}{h^2(R)} &\simeq & \frac{r^2}{R^2}\Bigg\{1+\frac{2 M}{r\Big[1-\frac{\omega^2_p}{\omega^2_0}\left(1+\frac{ \tilde B^2_0 }{1-\tilde \omega_{\varphi}^2}\right)\Big]}\\\notag
&-&\frac{2 M}{R\Big[1-\frac{\omega^2_p}{\omega^2_0}\left(1+\frac{ \tilde B^2_0 }{1-\tilde \omega_{\varphi}^2}\right)\Big]}\\ \nonumber
&-&\frac{Q^2}{r^2\Big[1-\frac{\omega^2_p}{\omega^2_0}\left(1+\frac{ \tilde B^2_0 }{1-\tilde \omega_{\varphi}^2}\right)\Big]} \\
&+&\frac{Q^2}{R^2\Big[1-\frac{\omega^2_p}{\omega^2_0}\left(1+\frac{ \tilde B^2_0 }{1-\tilde \omega_{\varphi}^2}\right)\Big]}\Bigg\}.
\label{eq:hrhR}
\end{eqnarray}
Using Eqs.~(\ref{eq:lastdef}) and (\ref{eq:hrhR}), we obtain 
\begin{eqnarray}\nonumber
    \hat{\alpha}&\simeq\frac{2M}{R}\bigg[1+\frac{1}{1-\frac{\omega_{\text{p}}^2}{\omega_0^2}\Big(1+\frac{ \tilde{B}^2_0 }{1-\tilde{\omega}_{\varphi}^2}\Big) }\bigg]-\\ 
    &-\frac{\pi Q^2}{4R^2}\bigg[1+\frac{2}{1-\frac{\omega_{\text{p}}^2}{\omega_0^2}\Big(1+\frac{ \tilde{B}^2_0 }{1-\tilde{\omega}_{\varphi}^2}\Big) }\bigg].
\end{eqnarray}
If $b$ denotes the impact parameter of the light ray, than for $R \simeq b$ and considering a uniform plasma $\omega^2_{p}=\text{const.}$, we obtain
\begin{eqnarray}\nonumber
    \hat{\alpha}(b)&\simeq\frac{2M}{b}\bigg[1+\frac{1}{1-\frac{\omega_{\text{p}}^2}{\omega_0^2}\Big(1+\frac{ \tilde{B}^2_0 }{1-\tilde{\omega}_{\varphi}^2}\Big) }\bigg]-\\ 
    &-\frac{\pi Q^2}{4b^2}\bigg[1+\frac{2}{1-\frac{\omega_{\text{p}}^2}{\omega_0^2}\Big(1+\frac{ \tilde{B}^2_0 }{1-\tilde{\omega}_{\varphi}^2}\Big) }\bigg]. \label{eq:lastdefuni}
\end{eqnarray}

Now, we can determine an expansion of the deflection angle expression for small values of the plasma frequency ($\omega^2_{p}/\omega^2_{0}\ll1$):
\begin{eqnarray}\nonumber
    \hat{\alpha}(b)&\simeq&\frac{2M}{b}\bigg[1+\frac{1}{1-\frac{\omega_{\text{p}}^2}{\omega_0^2}\Big(1+\frac{ \tilde{B}^2_0 }{1-\tilde{\omega}_{\varphi}^2}\Big) }\bigg]\\\notag
    &-&\frac{\pi Q^2}{4b^2}\bigg[1+\frac{2}{1-\frac{\omega_{\text{p}}^2}{\omega_0^2}\Big(1+\frac{ \tilde{B}^2_0 }{1-\tilde{\omega}_{\varphi}^2}\Big) }\bigg] \\\notag
    & \simeq& \Big(\frac{4 M}{b}-\frac{3 \pi Q^2}{4b^2}\Big)+\Big(\frac{2 M}{b}-\frac{\pi Q^2}{2b^2}\Big)\frac{\omega^2_{p}}{\omega^2_0}\\
    &+&\Big(\frac{2 M}{b}-\frac{\pi Q^2}{2b^2}\Big)\frac{\omega^2_{p}}{\omega^2_0} \tilde B^2_0(1+\tilde \omega_{\varphi}^2), \label{eq:lastdefuniseries}
\end{eqnarray}
where the first term corresponds to the gravitational field, the second term is the plasma contribution and the last term is due to axion-plasma fluid.
In Fig.~\ref{plot:defbuni} we depict the deflection angle versus the impact parameter $b$ of the photons for fixed values of the parameters $Q^2, \tilde{B}^2_0$, $\tilde \omega_{\varphi}^2$ and $\omega^2_{p}/\omega^2_0$. 
Figure~\ref{plot:defuni} represents the dependence of the deflection angle on the charge of BH and the axion fluid parameters for fixed values of impact parameter $b$ and of $\omega^2_{p}/\omega^2_0$.

\section{Deflection of light and relativistic massive particles using the Gauss-Bonnet theorem}\label{sec:massive}

In this section, we shall consider the problem of computing the deflection angle for relativistic massive particles. For the same reason,  let us consider the physical spacetime metric to be described by a BH surrounded by plasma medium. One can get the optical metric for investigating the deflection of light using $ds^2=0$, in the equatorial plane, this yields
\begin{equation}
dt^2=\frac{dr^2}{f^2(r)}+\frac{r^2 d\phi^2}{f(r)}.\label{38}
\end{equation}

We can use the following Gauss-Bonnet theorem (GBT) to study the deflection of light and that of relativistic massive particles. \\
\begin{figure}
 \begin{center}
   \includegraphics[scale=0.23]{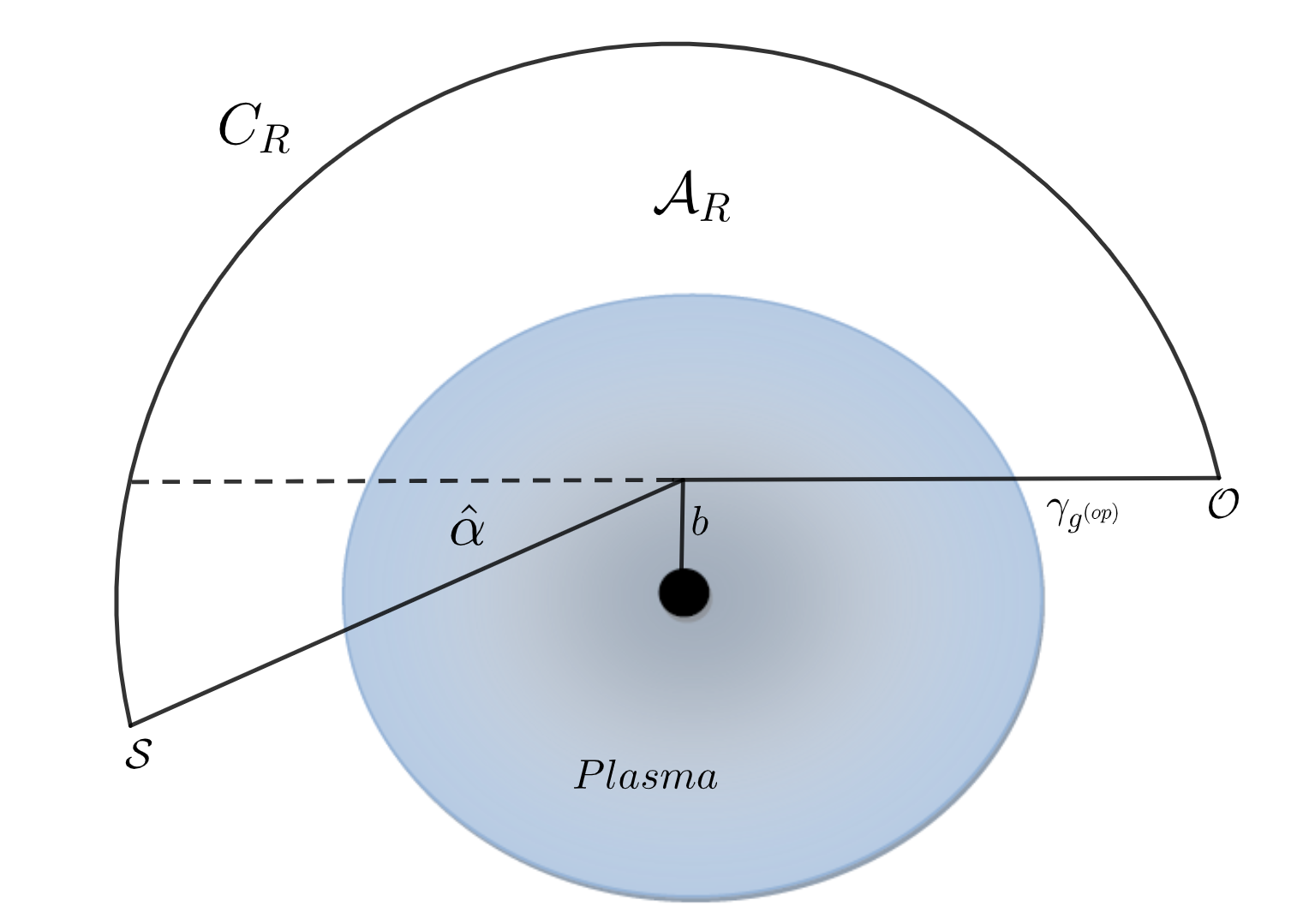}
  \end{center}
\caption{Schematic representation of the optical geometry of the BH surrounded by plasma. At the points $\mathcal{O}$ and $\mathcal{S}$, the interior angles satisfy the condition $\theta _{\mathcal{O}}+\theta _{\mathcal{S}}\rightarrow \pi $.}    \label{FigGB}
\end{figure}
\noindent \textbf{Theorem}: \textit{Let $\mathcal{A}_{R}$  be a non-singular domain with boundaries $\partial 
\mathcal{A}_{R}=\gamma_{g^{(op)}}\cup C_{R}$ of an oriented two-dimensional optical surface $S$ (see Fig~\ref{FigGB}) with the optical metric $g^{(op)}$. Let $K$ and $\kappa $ be the Gaussian optical
curvature and the geodesic curvature, respectively. Then, the GBT in terms of the above construction is written as follows} \cite{Gibbons:2008rj}
\begin{equation}
\int\limits_{\mathcal{A}_{R}}K\,dS+\oint\limits_{\partial \mathcal{%
A}_{R}}\kappa \,dt+\sum_{k}\delta _{k}=2\pi \chi (\mathcal{A}_{R}).
\label{10}
\end{equation}

In the GBT we have the optical surface element noted as $dS$ and the exterior angle at the corresponding $k^{th}$ vertex noted by $\delta_{k}$. It is rather interesting to see that the domain of integration is outside the light ray in the $(r,\phi)$ optical plane having the Euler characteristic number one, i.e., $\chi (\mathcal{A}_{R})=1$. Moreover if we introduce a smooth curve via $\gamma:=\{t\}\to \mathcal{A}_{R}$, we can compute the geodesic optical curvature  using the definition \cite{Gibbons:2008rj}
\begin{equation}
\kappa =g^{(op)}\,\left( \nabla _{\dot{\gamma}}\dot{\gamma},\ddot{\gamma}%
\right). 
\end{equation}
To simplify our calculations we are going to assume the unit speed condition given by $g^{(op)}(\dot{\gamma},\dot{\gamma})=1$, with $\ddot{\gamma}$ which stands for the unit acceleration vector. In the physical geometry the observer is located far away from the BH, hence by the same analogy, we can consider very large radial distance $r \equiv R\rightarrow \infty $, in such a limit, therefore we can express them in terms of the interior angles using $\theta _{\mathcal{O}}=\pi-\delta_{\mathcal{O}}$ and $\theta _{\mathcal{S}}=\pi-\delta_{\mathcal{S}}$. One can see that by construction, the two jump angles become $\pi/2$ (the jump angle at the source $\mathcal{S}$ and observer $\mathcal{O}$, respectively)  and should satisfy the condition $\theta _{\mathcal{O}}+\theta _{\mathcal{S}}\rightarrow \pi $ \cite{Gibbons:2008rj}. As we know, the geodesic optical curvature for the light ray vanishes, that is  $\kappa (\gamma_{g^{(op)}})=0$. From the GBT it follows that \cite{Gibbons:2008rj}
\begin{equation}
\lim_{R\to\infty }\int_{0}^{\pi+\hat{\alpha}}\left[\kappa \frac{d t}{d \phi}\right]_{C_R} d \phi=\pi-\lim_{R\to\infty }\int\limits_{\mathcal{A}_{R}}K\,dS.
\end{equation}

The nonzero contribution of the geodesic curvature for the curve $C_{R}$ is found by using  \cite{Gibbons:2008rj}
\begin{equation}
\kappa (C_{R})=|\nabla _{\dot{C}_{R}}\dot{C}_{R}|,
\end{equation}
and one can show the condition 
\begin{eqnarray}
\lim_{R\rightarrow \infty }\left(\kappa(C_R)\frac{dt}{d\phi}\right)=1.
\end{eqnarray}

Note that, the above condition is true only for asymptotically flat spacetimes. For static spacetimes in the presence of an optical medium, it has been shown that the optical metric and the spatial part of the spacetime metric are related by~\cite{Crisnejo:2018uyn}
\begin{eqnarray}\label{refin}
    g_{ij}^{op}=-\frac{n^2}{f(r)}g_{ij},\label{44}
\end{eqnarray}
where $i,j=1,2$. In other words, the spatial projections of the
light rays on the slices with $t = constant$ that solve
Hamilton’s equations are also spacelike geodesics of
the optical metric.
Let us also note that to compute the Gaussion optical curvature $K$, we can use the relation $K=R/2$, where $R$ is the Ricci scalar for the optical metric.

\subsection{Plasma medium with $\omega_{\text{p}}^2=\text{const.}$}
The simplest model corresponds to a medium with a uniform distribution of plasma. That is, the refractive index is given by
\begin{equation}
n^2(r)\simeq 1-\frac{\omega_{\text{p}}^2}{\omega_0^2}f(r) \left(1+\frac{ \tilde{B}^2_0 }{1-\tilde{\omega}_{\varphi}^2}\right).\label{45}
\end{equation}
Usings Eqs. (\ref{38}), (\ref{44}) and (\ref{45}), we can recast the optical metric of the BH metric surrounded by the plasma as
\begin{eqnarray}\nonumber
dt^{2}&=&\left[1-\frac{\omega_{\text{p}}^2}{\omega_0^2}f(r) \left(1+\frac{ \tilde{B}^2_0 }{1-\tilde{\omega}_{\varphi}^2}\right)\right]\nonumber\\&&\times\Big[ \frac{dr^2}{f(r)^2} +\frac{r^2}{f(r)}d\phi^2\Big].
\end{eqnarray}
From this, we can compute the Gaussian optical curvature, and after considering series expansion around $M/b$ and $Q^2/b^2$, we obtain in leading order terms
\begin{eqnarray}\nonumber
K\simeq -\frac{M\left[2-\frac{\omega_{\text{p}}^2}{\omega_0^2}\left(1+\frac{ \tilde{B}^2_0 }{1-\tilde{\omega}_{\varphi}^2}\right) \right]}{r^3\,\left[1-\frac{\omega_{\text{p}}^2}{\omega_0^2}\left(1+\frac{ \tilde{B}^2_0 }{1-\tilde{\omega}_{\varphi}^2}\right) \right]^2}\\
+\frac{Q^2\left[3-\frac{\omega_{\text{p}}^2}{\omega_0^2}\left(1+\frac{ \tilde{B}^2_0 }{1-\tilde{\omega}_{\varphi}^2}\right) \right]}{r^4\,\left[1-\frac{\omega_{\text{p}}^2}{\omega_0^2}\left(1+\frac{ \tilde{B}^2_0 }{1-\tilde{\omega}_{\varphi}^2}\right) \right]^2}.
\end{eqnarray}
From the GBT, for the deflection angle we have
\begin{equation}
\hat{\alpha}=-\int\limits_{0}^{\pi }\int\limits_{\frac{b}{\sin \varphi }%
}^{\infty }K dS,
\end{equation}
where we also need the expression for the surface element approximated as
\begin{equation}
dS \simeq r \bigg[1-\frac{\omega_{\text{p}}^2}{\omega_0^2}\bigg(1+\frac{ \tilde{B}^2_0 }{1-\tilde{\omega}_{\varphi}^2}\bigg) \bigg] dr d\phi.
\end{equation}
Solving the last integral is not difficult, hence we obtain 
\begin{eqnarray}\nonumber
\hat{\alpha} \simeq \frac{2M}{b}\bigg[1+\frac{1}{1-\frac{\omega_{\text{p}}^2}{\omega_0^2}\Big(1+\frac{ \tilde{B}^2_0 }{1-\tilde{\omega}_{\varphi}^2}\Big) }\bigg]\\
-\frac{\pi Q^2}{4b^2}\bigg[1+\frac{2}{1-\frac{\omega_{\text{p}}^2}{\omega_0^2}\Big(1+\frac{ \tilde{B}^2_0 }{1-\tilde{\omega}_{\varphi}^2}\Big) }\bigg]
\end{eqnarray}
As expected, this result coincides with the expression for the deflection angle obtained by the standard geodesic methods given by Eq.~(\ref{eq:lastdefuniseries}). In addition, this result generalizes the deflection angle obtained in Ref. \cite{Crisnejo:2018uyn}. 

\subsection{Plasma medium with $\omega_{\text{p}}^2(r)=z_0/r^q$}
For this particular model, the refractive index is given by
\begin{equation}
n^2(r)\simeq 1-\frac{z_0}{r^q \omega_0^2}f(r) \bigg(1+\frac{ \tilde{B}^2_0 }{1-\tilde{\omega}_{\varphi}^2}\bigg).
\end{equation}
In our particular case, we can recast the BH metric surrounded by plasma as
\begin{equation}
dt^{2}=\Big[1-\frac{z_0}{r^q \omega_0^2}f(r) \Big(1+\frac{ \tilde{B}^2_0 }{1-\tilde{\omega}_{\varphi}^2}\Big)\Big]\Big[ \frac{dr^2}{f(r)^2} \\
+\frac{r^2}{f(r)}d\phi^2\Big].
\end{equation}
Performing series expansions, for the Gaussian optical curvature we obtain in leading terms 
\begin{equation}
K\simeq -\frac{2M}{r^3}+\frac{4 Q^2}{r^4}+\frac{q^2 z_0 }{2 r^{q+2} \omega_0^2}\Big(1+\frac{ \tilde{B}^2_0 }{1-\tilde{\omega}_{\varphi}^2}\Big).
\end{equation}
Hence, the deflection angle is found to be
\begin{equation}
\hat{\alpha}=-\int\limits_{0}^{\pi }\int\limits_{\frac{b}{\sin \varphi }%
}^{\infty }\Big[ -\frac{2M}{r^3}+\frac{4 Q^2}{r^4}+\frac{q^2 z_0 }{2 r^{q+2} \omega_0^2}\Big(1+\frac{ \tilde{B}^2_0 }{1-\tilde{\omega}_{\varphi}^2}\Big) \Big]\mathrm{d}S. 
\end{equation}
Finally, solving this integral we find
\begin{equation}
\hat{\alpha} \simeq \frac{4M}{b}-\frac{\pi Q^2}{b^2}-\frac{z_0 \sqrt{\pi} \Gamma(\frac{q+1}{2})}{\omega_0^2\, b^q\, \Gamma(\frac{q}{2})}\Big(1+\frac{ \tilde{B}^2_0 }{1-\tilde{\omega}_{\varphi}^2}\Big).
\end{equation}
Again, the last expression generalizes the result found in Ref.~\cite{Crisnejo:2018uyn}. 

\subsection{Plasma medium with $\omega_{\text{p}}^2(r)=b_0\, e^{-r/r_0}$}
In another example we consider a plasma medium with a refractive index containing an exponentially decaying term, given by \cite{Huang:2018rfn}
\begin{equation}
n^2(r)\simeq 1-\frac{b_0e^{-r/r_0}}{ \omega_0^2}f(r) \Big(1+\frac{ \tilde{B}^2_0 }{1-\tilde{\omega}_{\varphi}^2}\Big).
\end{equation}
Here for the optical metric we obtain
\begin{equation}
dt^{2}= \Big[1-\frac{b_0e^{-r/r_0}}{ \omega_0^2}f(r) \Big(1+\frac{ \tilde{B}^2_0 }{1-\tilde{\omega}_{\varphi}^2}\Big)\Big]\Big[ \frac{dr^2}{f(r)^2} \\
+\frac{r^2}{f(r)}d\phi^2\Big].
\end{equation}
For the Gaussian optical curvature we find 
\begin{equation}
K\simeq -\frac{2M}{r^3}+\frac{4 Q^2}{r^4}+\frac{ b_0\,e^{-r/r_0} (r-r_0) }{2\,r\,r_0^2\,\omega_0^2}\bigg(1+\frac{ \tilde{B}^2_0 }{1-\tilde{\omega}_{\varphi}^2}\bigg).
\end{equation}
The deflection angle is found
\begin{equation}
\hat{\alpha} \simeq \frac{4M}{b}-\frac{\pi Q^2}{b^2}-\frac{b \,b_0 K_0(b/r_0)}{r_0 \omega_0^2}\bigg(1+\frac{ \tilde{B}^2_0 }{1-\tilde{\omega}_{\varphi}^2}\bigg),
\end{equation}
where $K_0$ is the zeroth order modified Bessel function of the second kind. One can see that by setting $\tilde{B}_0^2=0$,  we recover the result derived in~\cite{Crisnejo:2018uyn}.

\subsection{Deflection of relativistic massive particles}
 We shall focus on the deflection of relativistic massive particles in the presence of axion-plasmon medium. To find the deflection angle for massive particles we proceed as follows. First, by following the Refs. \cite{Crisnejo:2018uyn,Crisnejo:2018ppm,Crisnejo:2019ril}, one can incorporate the refractive index of the medium in the optical metric, hence we can write
\begin{equation}
dt^2 \to d\sigma^2=n^2(r)\left(\frac{dr^2}{f^2(r)}+\frac{r^2 d\phi^2}{f(r)}\right),
\end{equation}
in the last equation we have used Eq. (\ref{refin}) to introduce the refractive index in the metric. 
Secondly, we use the correspondence between the motion of photons in plasma and massive particles i.e. we can identify the rest mass of the particle with the frequency of the plasma, and the energy of the particle with the photon frequency, hence (with $\hbar=1$)
\begin{eqnarray}
\omega_{\text{p}} \longrightarrow m_0, \,\,\,\omega_0 \longrightarrow E.
\end{eqnarray}
For the refractive index we can write 
\begin{equation}
n^2(r)\simeq 1-\frac{m_0^2}{E^2_{\infty}}f(r) \bigg(1+\frac{ \tilde{B}^2_0 }{1-\tilde{\omega}_{\varphi}^2}\bigg).\label{62}
\end{equation}
The relativistic particle with velocity $v$ has energy 
\begin{equation}
E_{\infty}=\frac{m_0}{\sqrt{1-v^2}},\label{63}
\end{equation}
as measured far away from the observer at spatial infinity. In addition, let us assume that the particle has an angular momentum given by
\begin{equation}
J=\frac{m_0 v\, b}{\sqrt{1-v^2}},\label{64}
\end{equation}
here $b$ is an impact parameter of the massive particle. Combining Eqs. (\ref{62}), (\ref{63}) and (\ref{64}), we find
\begin{equation}
n^2(r)\simeq 1-(1-v^2)f(r) \bigg(1+\frac{ \tilde{B}^2_0 }{1-\tilde{\omega}_{\phi}^2}\bigg)
\end{equation}
That is, we can recast the purely optical metric in presence of axion-plasmon medium in the equatorial plane as
\begin{equation}\label{65}
d\sigma^{2}=\Big[1-(1-v^2)f(r) \Big(1+\frac{ \tilde{B}^2_0 }{1-\tilde{\omega}_{\varphi}^2}\Big)\Big]\Big[ \frac{dr^2}{f^2(r)}
+\frac{r^2d\phi^2}{f(r)}\Big].
\end{equation}
The Gaussian optical curvature from the metric~(\ref{65}) is computed after we perform series expansions around $M/b$ and $Q^2/b^2$, we get
\begin{eqnarray}\nonumber
    K\simeq& -\frac{M \left[1+v^2-(\frac{ \tilde{B}^2_0 }{1-\tilde{\omega}_{\varphi}^2})(1-v^2)   \right]}{r^3\left[v^2-(\frac{ \tilde{B}^2_0 }{1-\tilde{\omega}_{\varphi}^2})(1-v^2)   \right]^2}\\
    &+\frac{ Q^2 \left[2+v^2-(\frac{ \tilde{B}^2_0 }{1-\tilde{\omega}_{\varphi}^2})(1-v^2)   \right]}{r^4\left[v^2-(\frac{ \tilde{B}^2_0 }{1-\tilde{\omega}_{\varphi}^2})(1-v^2)   \right]^2},
\end{eqnarray}
in leading order terms. 
Notice an interesting consequence  in the last equation in the limit $v \to 0$ and $\tilde{B}_0 \to 0$, yielding an apparent singularity in $K$. This apparent singularity shows that, we need to restrict our analyses only for relativistic particles, namely, the particles speed belongs to the following interval $0<v \leq 1$ (with $c=1$), while this also suggests that for nonrelativistic motions one must develop a different or more general setup. For the geodesic deviation having large radial coordinate $R$, yields
\begin{eqnarray}
\lim_{R\rightarrow \infty }\kappa (C_{R}) &\rightarrow &\frac{1}{\left(v^2-(\frac{ \tilde{B}^2_0 }{1-\tilde{\omega}_{\varphi}^2})(1-v^2)  \right) \,R}.
\end{eqnarray}
Using the metric~(\ref{65}), for $r=R$ held constant we find that
\begin{eqnarray}
\lim_{R\rightarrow \infty }d\sigma \rightarrow \left(v^2-(\frac{ \tilde{B}^2_0 }{1-\tilde{\omega}_{\varphi}^2})(1-v^2)\right) \,R\,d\phi.
\end{eqnarray}
This leads to the expected condition
\begin{eqnarray}
\lim_{R\rightarrow \infty }\left(\kappa(C_R)\frac{d\sigma}{d\phi}\right)=1.
\end{eqnarray}
To compute the deflection angle we need to use 
\begin{equation}
\hat{\alpha}=-\int\limits_{0}^{\pi }\int\limits_{\frac{\mathsf{b}}{\sin \varphi }%
}^{\infty }K dS. 
\end{equation}
where we need to integrate over the optical domain with the approximated surface element 
\begin{equation}
dS \simeq r\left(v^2-(\frac{ \tilde{B}^2_0 }{1-\tilde{\omega}_{\varphi}^2})(1-v^2)\right)  dr d\phi.
\end{equation}
Finally, evaluating this integral we find
\begin{eqnarray}\nonumber
\hat{\alpha} &\simeq \frac{2M}{b}\left[1+\frac{1}{\left(v^2-(\frac{ \tilde{B}^2_0 }{1-\tilde{\omega}_{\varphi}^2})(1-v^2)\right) }\right]-\\
&-\frac{\pi Q^2}{4b^2}\left[1+\frac{2}{\left(v^2-(\frac{ \tilde{B}^2_0 }{1-\tilde{\omega}_{\varphi}^2})(1-v^2)\right) }\right].
\end{eqnarray}
If we set $\tilde{B}_0^2=0$, we obtain the result earlier found in Ref.~\cite{Crisnejo:2018uyn}. Furthermore if we make the identification, 
\begin{eqnarray}
v^2 \longrightarrow 1-\frac{\omega_{\text{p}}^2}{\omega_0^2},
\end{eqnarray}
we obtain the same result as previously derived in the case of the deflection of light given in Eq. (\ref{eq:lastdefuni}).    

\section{Axion-plasmon effect on the Einstein Rings in the weak field}\label{sec:ring}
Let us now turn our attention and focus on the observational relevance of our results for the axion-plasmon model. Toward this purpose we shall use the expression for the deflection angles to estimate the size of the Einstein rings using three plasma models. Furthermore we can adopt the following setup:  The BH, or the the lens $L$, is located between the source $S$ and the observer $O$, and both $S$ and $O$ are located in the asymptotically flat region, i.e.,  at the distances much larger than the BH size. As we will see below, the Einstein rings can be formed due to the gravitational field of a BH when the source, lens and observer are perfectly aligned. In general, by construction, we can relate the observational angular coordinates, or the image position $\theta$, the source position $\beta$ and the light deflection angle $\alpha$ using the Ohanian lens equation~\cite{Bozza:2008ev}
\begin{equation}\label{LensEq}
    \arcsin\left(\frac{D_{OL}}{D_{LS}}\sin{\theta}\right)-\arcsin{\left(\frac{D_{OS}}{D_{LS}}\sin{\beta}\right)}=\hat{\alpha}(\theta)-\theta,
\end{equation}
where $D_{OL}$ represents the observer--lens distance, $D_{OS}$ represents the observer--source distance and $D_{LS}$ is the distance from the lens to the source (see corresponding figure in~\cite{Bozza:2008ev}). In the above equation the deflection angle $\hat{\alpha}$ is expressed in terms of $\theta$ using the relation for the impact parameter $b=D_{OL}\sin{\theta}$. In the literature, however, there are more general solutions of the lens equation (\ref{LensEq}), for example we can use \cite{Bozza:2008ev}
\begin{equation}\label{ImagePositions}
    D_{OS}\tan\beta=\frac{D_{OL}\sin\theta-D_{LS}\sin(\hat{\alpha}-\theta)}{\cos(\hat{\alpha}-\theta)}
\end{equation}
In the general case, one can show that the values of $\theta$, which are solutions to the above equations, and provide information about the positions of the weak field images. In the weak deflection approximation, both equations yield~\cite{Bozza:2008ev}
\begin{eqnarray}\label{EinsteinRing}
    \beta=\theta-\frac{D_{LS}}{D_{OS}}\hat{\alpha}.
\end{eqnarray}
For the Einstein ring to form, we need to consider the special case having $\beta=0$, i.e., the source $S$ lies on the optical axis. It is easily seen that in the weak deflection limit ($\hat{\alpha}\ll1, \beta\ll1$), we can use the last equation to compute the angular radius of the Einstein ring as follows
\begin{eqnarray}\label{EinsteinRing1}
    \theta_{E}\simeq\frac{D_{LS}}{D_{OS}}\hat{\alpha}(b).
\end{eqnarray}
Here we have used the relation $D_{OS}=D_{OL}+D_{LS}$ provided the angular source position is $\beta=0$. 

In order to see the axion-plasmon effect on the Einstein rings, let us take as an example the BH with mass $M=4.31\times 10^{6}M_{\odot}$ located at our galactic center Sgr A$^{*}$, with an observer locate at the distance $D_{OL}=8.33$ kpc from the Sgr A$^{*}$ (lens). Furthermore we shall assume the following $D_{LS}=D_{OL}/2$ meaning that $D_{OS}=3D_{OL}/2$. Let us take, as a first example the case of homogeneous plasma, to obtain the angular scale in the celestial sky. To first order of approximation in the deflection angle and using the relation $b=D_{OL}\sin{\theta}\simeq D_{OL}\theta$, the bending angle in the hypothesis of small angles is
\begin{eqnarray}\nonumber
    \hat{\alpha}(\theta)&\simeq\frac{2M}{D_{OL}\theta}\Bigg(1+\frac{1}{1-\frac{\omega_{\text{p}}^2}{\omega_0^2}\left(1+\frac{ \tilde{B}^2_0 }{1-\tilde{\omega}_{\varphi}^2}\right)}\Bigg)\\
    &-\frac{\pi Q^2}{4(D_{OL}\theta)^2}\Bigg(1+\frac{2}{1-\frac{\omega_{\text{p}}^2}{\omega_0^2}\left(1+\frac{ \tilde{B}^2_0 }{1-\tilde{\omega}_{\varphi}^2}\right)}\Bigg).
\end{eqnarray}%
\begin{figure}
\includegraphics[scale=0.45]{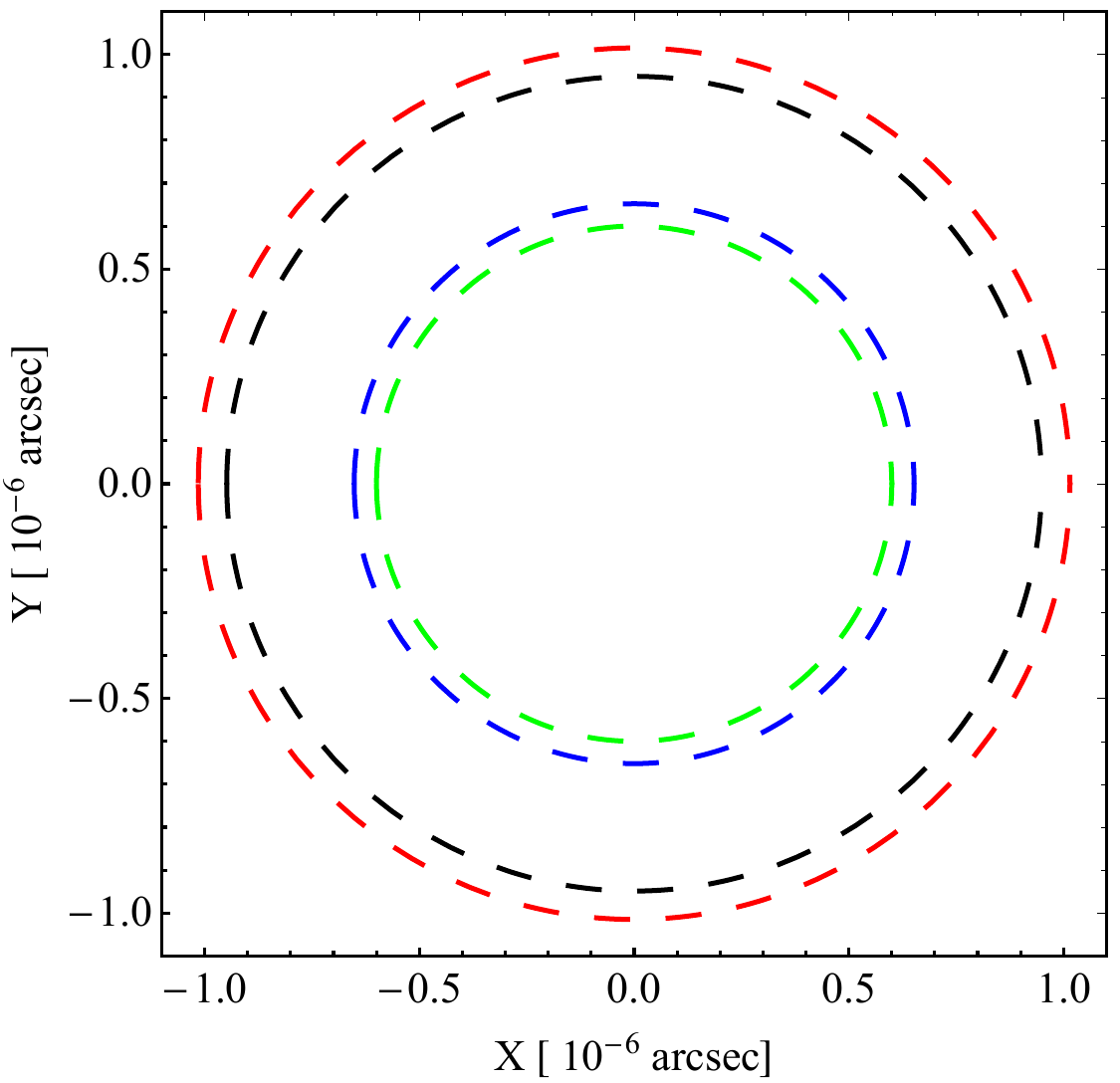}
\caption{The positions of the weak-field Einstein rings by the BH surrounded by a homogeneous plasma (red curve), Reissner–Nordström BH (black curve) and inhomogeneous power law plasma (blue curve) and inhomogeneous exponential law plasma (green curve). We have set $\omega_{\text{p}}^2/\omega_0^2=0.3$, $z_0/\omega_0^2=0.3 [M]$ , $Q^2=0.3[M]$, $\tilde{B}_0^2=\tilde{\omega}^2_{\varphi}=0.5$, $b \sim 10 r_0$ and $b_0/\omega_0^2=0.1$.  Here $X$ and $Y$ are the angular celestial coordinates in the observer's sky.\label{Fig8}}
\end{figure}%
Considering for instance the case $\omega_{\text{p}}^2/\omega_0^2=0.3$, $\tilde{B}_0^2=\tilde{\omega}^2_{\varphi}=0.5$, $Q^2=0.3[M]$ and the quantity Q is measured in units of the BH mass and using the approximation $M/D_{OL}\approx 2.48\times 10^{-11}$, we obtain $\vartheta_{E}\simeq 1.57 \, \text{arcsec}$ and $\vartheta_{E}\simeq 1.03 \times 10^{-6} \, \text{arcsec}$, which is larger than the corresponding value for  the Reissner–Nordström BH case $\vartheta_{E}^{\rm Sch}\simeq 1.18 \, \text{arcsec}$ and $\vartheta_{E}\simeq 0.90 \times 10^{-6} \, \text{arcsec}$ .  Although this is a small effect, in principle, there is a possibility for detecting this axion effect by observation of the rings. For nonzero axion dark matter parameters, we find that there is a larger size of the relativistic rings, compared to the Reissner–Nordström BH ring. Let us also consider the model $\omega^2_p(r)=z_0/r^q$, with $q=1$, then 
\begin{eqnarray}
    \hat{\alpha}(\theta)\simeq\frac{4M-\frac{z_0}{\omega_0^2}\left(1+\frac{ \tilde{B}^2_0 }{1-\tilde{\omega}_{\varphi}^2}\right)}{D_{OL}\theta}-\frac{\pi Q^2}{(D_{OL}\theta)^2}.
\end{eqnarray}

\begin{figure*}
\includegraphics[scale=0.7]{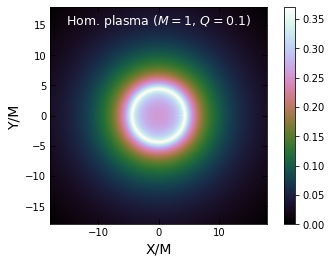}
\includegraphics[scale=0.7]{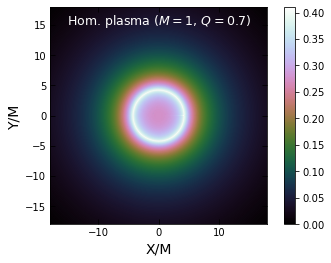}
\includegraphics[scale=0.7]{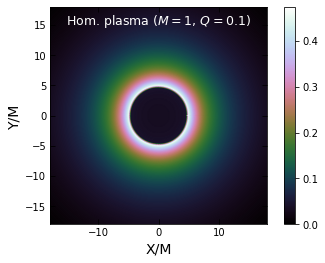}
\includegraphics[scale=0.7]{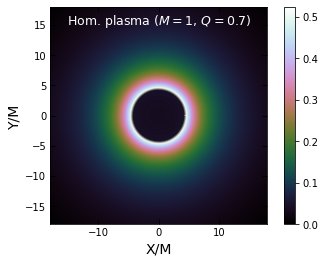}
\caption{Shadow images and the corresponding intensities (upper panel for rest gas) for a charged BH in the presence of homogeneous plasma (lower panel for infalling gas). We have set $\omega_{\text{p}}^2/\omega_0^2=0.3$, $\tilde{B}_0^2=\tilde{\omega}^2_{\varphi}=0.5$.} \label{shadowimages1}
\end{figure*} 

\begin{figure*}
\includegraphics[scale=0.7]{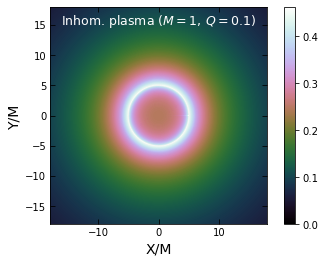}
\includegraphics[scale=0.7]{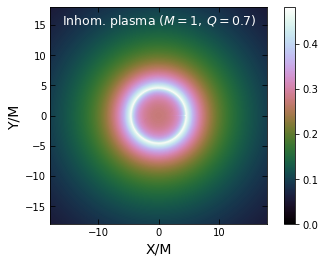}
\includegraphics[scale=0.7]{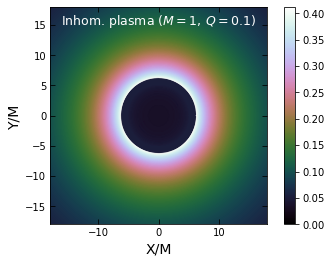}
\includegraphics[scale=0.7]{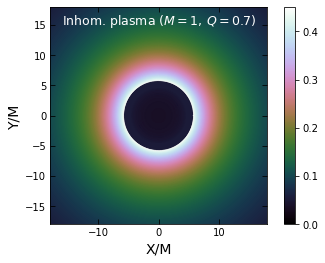}
\caption{Shadow images and the corresponding intensities (upper panel for rest gas) for a charged BH in the presence of inhomogeneous plasma  (lower panel for infalling gas).  We have set $z_0/\omega_0^2=0.3\, [M]$, $\tilde{B}_0^2=\tilde{\omega}^2_{\varphi}=0.5$, $q=1$. } 
\label{shadowimages2}
\end{figure*}

\begin{figure*}
\includegraphics[scale=0.7]{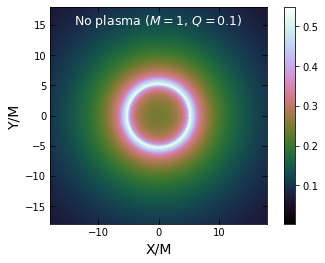}
\includegraphics[scale=0.7]{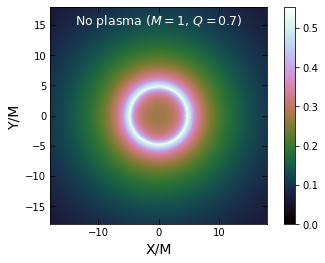}
\includegraphics[scale=0.7]{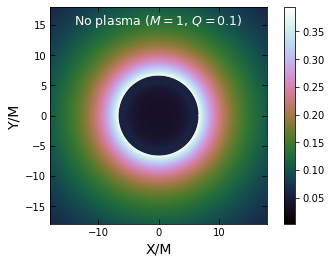}
\includegraphics[scale=0.7]{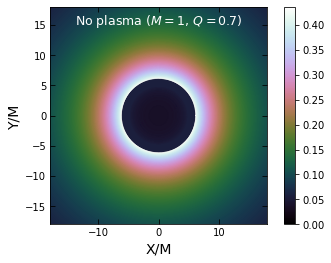}
\caption{Shadow images and the corresponding intensities  (upper panel for infalling gas). for a charged BH in the absence of plasma  (lower panel for infalling gas).  } \label{shadowimages3}
\end{figure*}  

Note that the quantity $z$ is measured in units of the BH mass. Taking for example, $z_0/\omega_0^2=0.3\, [M]$, $Q^2=0.3[M]$, $\tilde{B}_0^2=\tilde{\omega}^2_{\varphi}=0.5$, we obtain $\vartheta_{E}\simeq 1.09 \, \text{arcsec}$ and $\vartheta_{E}\simeq 0.425 \times 10^{-6} \, \text{arcsec}$, which is smaller compared to the Reissner–Nordström BH. This means that the deflection angle and the size of Einstein rings depends on the particular plasma frequency model. For the exponential plasma model a closed form for the Einstein ring is not possible to obtain, however, one can only approximate the numerical value. We can simplify the problem by assuming in the exponential model $b_0\,e^{-r/r_0}$ a scale radius of kpc orders, say $b \sim 10 \,r_0 $, yielding
\begin{equation}
\hat{\alpha}(\theta) \simeq \frac{4M}{D_{OL} \theta}-\frac{\pi Q^2}{(D_{OL} \theta)^2}-\frac{10\,b_0 K_0(10)}{\omega_0^2}\bigg(1+\frac{ \tilde{B}^2_0 }{1-\tilde{\omega}_{\varphi}^2}\bigg)
\end{equation}
Taking  $b_0/\omega_0^2 \sim 0.1$ along with $\tilde{B}_0^2=\tilde{\omega}^2_{\varphi}=0.5$, $Q^2=0.3[M]$, we find $\vartheta_{E}\simeq  0.48 \, \text{arcsec}$ and $\vartheta_{E}\simeq 0.36 \times 10^{-6} \, \text{arcsec}$, which is also smaller compared to the Reissner–Nordström BH.

\section{Shadow images in a plasma medium}
\subsection{Infalling gas in a plasma medium}\label{infalling}
In this section, we are going to study a simple accretion model which consists of an infalling gas onto the charged black hole surrounded by a different model of plasma, i.e., axion-plasmon medium. In particular, we are going to use the numerical technique known as Backward Raytracing in order to find the apparent shadow due to the infalling and radiation gas \cite{Falcke:1999pj,Bambi:2013nla,Bambi:2017khi,Saurabh:2020zqg,Jusufi:2020zln, Jusufi:2021lei,Shaikh:2018lcc}. That being said, let us define the specific intensity $I_{\nu 0}$ which should be observed at some large distance from the black hole and it is given by the expression \cite{Bambi:2013nla}
\begin{eqnarray}
    I_{obs}(\nu_{obs},X,Y) = \int_{\gamma}\mathrm{g}^3 j(\nu_{e})dl_{\text{prop}},\,
\end{eqnarray}
where $g=\nu_{obs}/\nu_e$ is the redshift factor and $\nu_e$ gives the photon
frequency which is measured in the rest-frame of the emitter. To calculate the total flux one can use the relation \cite{Bambi:2013nla,Nampalliwar:2020asd}
\begin{eqnarray}\label{flux}
    F_{obs}(X,Y) =\int_{\gamma} I_{obs}(\nu_{obs},X,Y) d\nu_{obs}.
\end{eqnarray}

The radiating gas is in a free fall so that its four-velocity components are given by \cite{Bambi:2013nla}
\begin{equation}
u_e^{\mu}=\left(\frac{1}{f(r)}, -\sqrt{1-f(r)}, 0, 0 0\right).
\end{equation}

In order to compute the total flux we also need to determine the relation between the radial and time components of the photon four-velocity which is given by the relation
\begin{equation}
    k^r= \pm k^t f(r)\,\sqrt{f(r)\bigg(\frac{1}{f(r)}-\frac{b^2}{r^2}-\frac{\omega_{\text{p}}^2(r)}{\omega_0^2} \bigg(1+\frac{\tilde{B}^2}{1-\tilde{\omega }_{\varphi }^2}\bigg)\bigg)}.
\end{equation}

The physical meaning of the signs $+(-)$ in the above equation is the following: The photon can either approach or recedes from the BH. Note that the impact parameter $b$ encodes the axion-plasmon effect and it reads
\begin{equation}
b =\frac{p_{\phi}}{\omega_0}= r\sqrt{ \frac{1}{f(r)} - \frac{\omega_{\text{p}}^2(r)}{\omega_0^2} \bigg(1+\frac{\tilde{B}^2}{1-\tilde{\omega }_{\varphi }^2}\bigg)}|_{r=R} \,.
\end{equation}
We can also use the redshift function $\mathrm{g}$ which can be calculated also by the relation \cite{Bambi:2013nla}
\begin{eqnarray}
   \mathrm{g} = \frac{k_{\alpha}u^{\alpha}_o}{k_{\beta}u^{\beta}_e},
\end{eqnarray}
In our accretion model, we shall apply one more assumption, namely we are going to use a monochromatic and a  $1/r^2$ radial profile for the specific emissivity given by the equation
\begin{eqnarray}
    j(\nu_{e}) \propto \frac{\delta(\nu_{e}-\nu_{\star})}{r^2},
\end{eqnarray}
in which $\delta$ is the Dirac delta function. 

\subsection{Rest gas in plasma medium}
Another interesting model to use is the rest spherical accretion model. In this model the redshift factor reads $\mathrm{g}=f(r)^{1/2}$. Assuming again the emission to be monochromatic and radial profile $r^{-2}$ with the specific intensity observed at some large distance
\begin{equation}
I_{obs}(\nu_{obs}) = \int_{\gamma} \frac{f(r)^{3/2}}{r^2} \sqrt{f(r)^{-1}+r^2 \left(\frac{d\phi}{dr}\right)^2}  dr 
\end{equation}
along with
\begin{eqnarray}
\frac{d\phi}{dr}=\pm \frac{1}{r \sqrt{f(r)\left(\frac{h^2(r)}{b^2}-1\right)}}.
\end{eqnarray}

\subsection{Rotating gas in plasma medium}
One can also assume a rotating and radiating gas around the black hole with the four-velocity components 
\begin{eqnarray}
u^\mu_e=u^{t}\Big(1,0,0,\Omega \Big),
\end{eqnarray}
where $u^{t}=(f(r)-r^2 \Omega^2)^{-1/2}$, and $\Omega=\sqrt{f'(r)/2r}$. Moreover one has to use the relation for the photon $\mathcal{H}=0$, and the redshift function $\mathrm{g}$ given by Eq. (86). In the present work, we assume that the specific emissivity is described by the radial law $r^{-2}$. Finally, once we express the proper length  and after we integrate the intensity over all the observed frequencies, that is, we can write \cite{Bambi:2013nla}
\begin{equation}\label{inten}
    F_{obs}(X,Y) \propto -\int_{\gamma} \frac{\mathrm{g}^3 k_t}{r^2k^r}dr.  
\end{equation}

We closely follow the numerical technique presented in \cite{Saurabh:2020zqg,Jusufi:2020zln, Jusufi:2021lei} and the resulting shadow images and the intensities of the charged black hole with the homogeneous plasma with axion-plasmon effects are depicted in Fig.~\ref{shadowimages1}. In this plot, $X$ and $Y$ are the angular celestial coordinates in the observer's sky.  As we can see in the case of uniform plasma medium  there is a difference in the intensities as well as the shadow radii, namely when the charge increases, the intensities increases, however, the size of the shadow radius decreases. For this specific case we have set $M=1$, and $Q=0.1$ and $Q=0.7$, along with  $\omega_{\text{p}}^2/\omega_0^2=0.3$, $\tilde{B}_0^2=\tilde{\omega}^2_{\varphi}=0.5$, yielding the photon sphere $r_{ph}/M=3.304281794$ and $r_{ph}/M=2.897241801$, respectively. For the location of the horizon we get $r_{+}/M=1.994987437$  and $r_{+}/M=1.714142843$. For the shadow radius we have $R_{sh}/M=4.587535892$ and $R_{sh}/M=4.215386740$, respectively. The fact that the shadow radius decreases with the increase of charge directly is related to the horizon size, which also decreases with the increase of charge. 

Next, the shadow images and the intensities of the charged black hole with the inhomogeneous plasma with axion-plasmon effects are depicted in Fig.~\ref{shadowimages2}. Again, we obtain similar results, with the increase of charge the intensities increase, and the shadow radius decreases. Compared to the case of homogeneous plasma, we see a very different optical appearance of the shadow. For this case we have also set $M=1$, along with $Q=0.1$ and $Q=0.7$. For the plasma model we take  $\omega_{\text{p}}^2=z_0/r^q$, with $q=1$, and $z_0/\omega_0^2=0.3$ along with $\tilde{B}_0^2=\tilde{\omega}^2_{\varphi}=0.5$. We get for the photon sphere $r_{ph}/M=3.028204716$ and $r_{ph}/M=2.662197042$, respectively. For the shadow radius we have $R_{sh}/M=5.010349436$ and $R_{sh}/M=4.549535862$, respectively. This shows that the shadow radius decreases with the increase of charge which is again linked to the decrease of the horizon with the increase of charge.

Finally, we present the shadow images and the intensities of the charged black hole in absence of the plasma medium  depicted in Fig.~\ref{shadowimages3}. With the increase of charge, the intensities increase, and the shadow radius decreases. We observe that the most dramatic optical appearance is obtained when we compare the images to the homogeneous plasma. In particular, we have set $M=1$, along with $Q=0.1$ and $Q=0.7$ and $\omega_{\text{p}}^2=0$. We get for the photon sphere $r_{ph}/M=2.993318452$ and $r_{ph}/M=2.626942767$, respectively. For the shadow radius we have $R_{sh}/M=5.187475262$ and $R_{sh}/M=4.720682333$, respectively. We thus found that for constant value of charges the shadow radius is larger in the case of no plasma, and smaller in the case of homogeneous plasma.  \\

This difference in the intensities as we get away from the black hole, is explained by the fact that the deflection angle of light is affected by plasma. Since the deflection angle increases for the uniform plasma, the intensity will be smaller at infinity since more photons will be captured by the BH. In the present work, we have integrated numerically from the photon sphere, although there is a small contribution, or practically a neglecting effect, coming from the region between the horizon and the photon sphere. 

\section{Constraints on BH shadow}
In this section, we perform constraints on BH shadow as done in \cite{Davlataliev:2023ckw}. We use the following $1\sigma$ and $2\sigma$ constraints on $R_{sh}/M$\cite{Vagnozzi_2023}:
\begin{align}\label{constraints}
    1\sigma:   \qquad   4.55\lesssim R_{sh}/M \lesssim 5.22,\\
    2\sigma:   \qquad   4.21\lesssim R_{sh}/M \lesssim 5.56.
\end{align}

\begin{figure}
    \centering
    \includegraphics[scale=0.45]{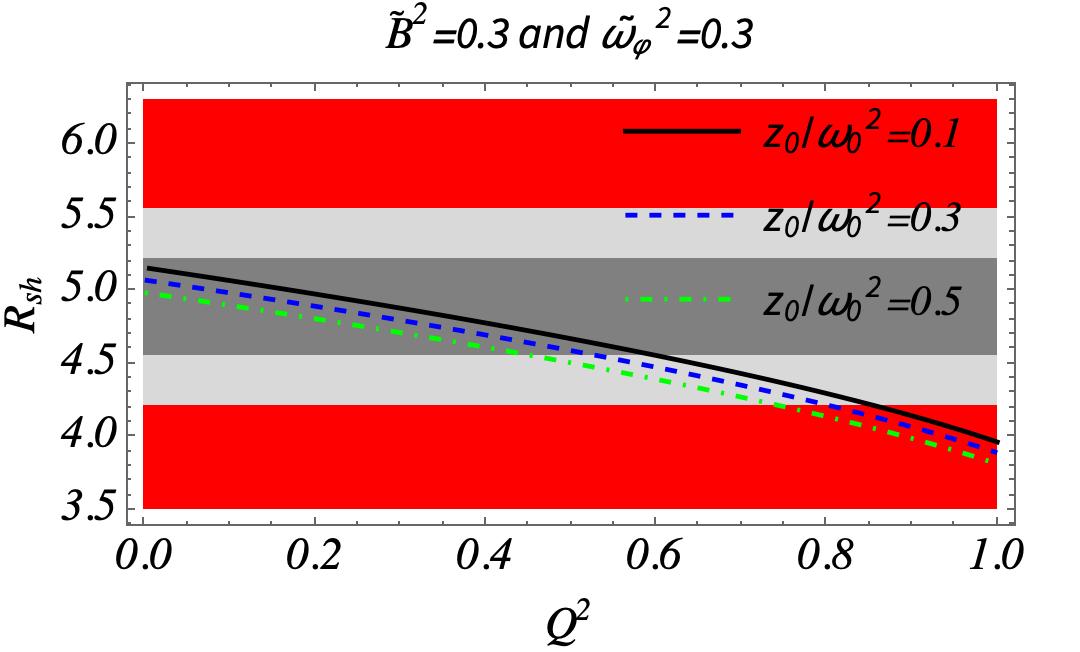}
    \caption{The shadow radius in units of the BH mass M, as a function of the normalized electric charge $Q/M$, as discussed in Sec. \ref{2.A}. The regions shaded in dark gray and light gray correspond to the $1\sigma$ and $2\sigma$ respectively, in agreement with the Event Horizon Telescope's horizon-scale image of Sgr A$^*$. The red regions are instead excluded by the same observations at more than $2\sigma$.}
    \label{fig:conRQ}
\end{figure}

These constraints are in good agreement with the observation for Sgr A$^{*}$'s shadow which was reported in Ref. \cite{2022ApJ...930L..17E}. The changes in the shadow radius depending on the electric charge are plotted in Fig. \ref{fig:conRQ} , illustrating the effects of the plasma frequency on the Reissner-Nordström spacetime. These plots are plotted within the constraints set by observations from the Event Horizon Telescope image of Sgr A$^*$ [Eq. (\ref{constraints})]. We can see that with an increase in electric charge, the shadow radius decreases. This phenomenon can be understood by examining how the electric charge influences the potential experienced by test particles. Moreover, as the plasma frequency increases, the shadow radius decreases.

\begin{figure}
    \centering
    \includegraphics[scale=0.57]{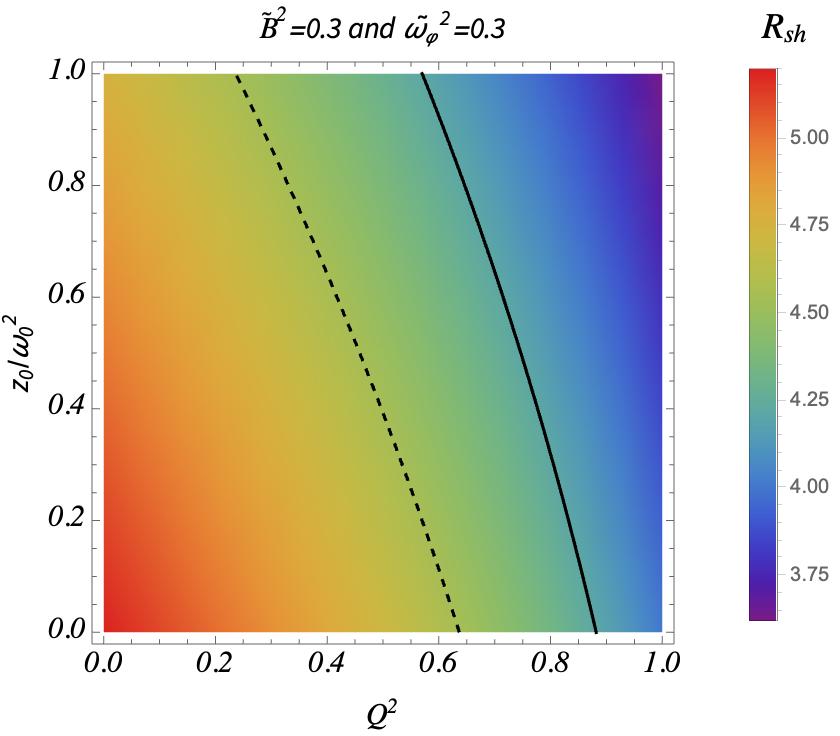}
    \caption{Constraint on the BH charge $Q^2$ and the plasma frequency $\frac{z_0}{\omega_0^2}$. The dashed and solid lines correspond to the minimum boundaries of $1\sigma$ and $2\sigma$ respectively.}
    \label{fig:conQz}
\end{figure}

\begin{figure}
    \centering
    \includegraphics[scale=0.57]{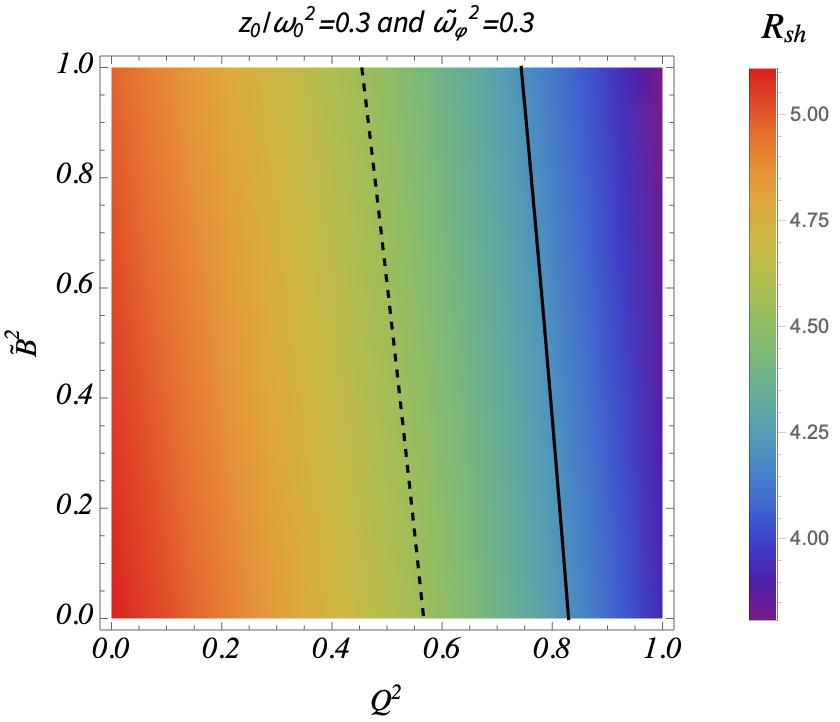}
    \caption{Constraint on the BH charge $Q^2$ and the magnetic field $\tilde{B}^2$. The dashed and solid lines correspond to the minimum boundaries of $1\sigma$ and $2\sigma$ respectively.}
    \label{fig:conQB}
\end{figure}

\begin{figure}
    \centering
    \includegraphics[scale=0.57]{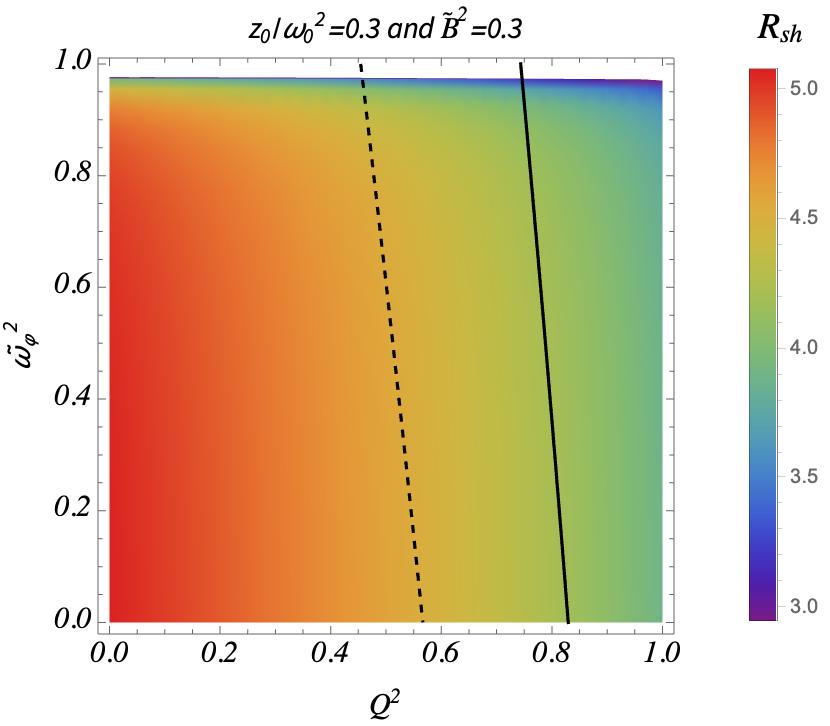}
    \caption{Constraint on the BH charge $Q^2$ and the axion frequency $\tilde{\omega}_\phi^2$. The dashed and solid lines correspond to the minimum boundaries of $1\sigma$ and $2\sigma$ respectively.}
    \label{fig:conQw}
\end{figure}

In Fig. \ref{fig:conQz},\ref{fig:conQB} and \ref{fig:conQw}, we investigate the effects of the plasma frequency ${z_0}/{\omega_0^2}$
, the magnetic field $\tilde{B}^2$ and the axion frequency $\tilde{\omega}_\phi^2$ on the radius of Reissner–Nordström BH shadow. The sets of values for  $\{{z_0}/{\omega_0^2}, Q^2\}$ , $\{\tilde{B}^2, Q^2\}$ and $\{\tilde{\omega}_\phi^2, Q^2\}$ in the left-hand side of dashed and solid lines in Fig. \ref{fig:conQz}, \ref{fig:conQB}, \ref{fig:conQw} correspond to the $1\sigma$ and $2\sigma$ respectively.

\section{Conclusions}
\label{Sec:conclusion}
In this work, we have investigated the axion-plasmon effect on the optical properties of the charged BH. In doing so, we present the modified equations of motion of photons around the charged black hole. 

We have investigated in details a number of observational effects such as the  black hole shadow, the gravitational deflection angle, Einstein rings and shadow images obtained by radially infalling gas on a black hole within different models of the plasma medium. It is found that the intensity of the electromagnetic radiation increases with the increase of charge and the size of the black hole shadow decreases with increase of the electric charge for a fixed axion-plasmon coupling values when observed from sufficiently large distances.  Finally, from the constraning analyses using the Sgr A$^{\star}$ data it is shown that the  parameters domain used in the present work are reasonable with the data of Sgr A$^{\star}$ black hole shadow. As the effect of plasma medium can be significant, in principle, this means that with the improvement of the precision of measuring the shadow of black holes we can detect the matter fields around black holes such as plasma medium.
\section*{Acknowledgements} This research is partly supported by Research Grant F-FA-2021-510 of the Uzbekistan Ministry for Innovative Development.

\bibliographystyle{apsrev4-1}
\bibliography{Shadow}

\end{document}